\documentclass{aa}
\makeatletter
\def\jobname{AGNFormer}
\makeatother
\usepackage{graphicx}
\usepackage{txfonts}
\usepackage{lipsum}
\usepackage{subcaption}         
\usepackage{lscape}             
\usepackage{placeins}           

\usepackage{txfonts}
\newcommand{\lyman}{Ly$\alpha$}
\newcommand{\hb}{H$\beta$}
\newcommand{\ha}{H$\alpha$}
\newcommand{\hg}{H$\gamma$}
\newcommand{\hd}{H$\delta$}
\newcommand{\he}{H$\epsilon$}
\newcommand{\mg}{Mg\,\textsc{ii}}
\newcommand{\civ}{C\,\textsc{iv}}
\newcommand{\oi}{O\,\textsc{i}}
\newcommand{\siiv}{Si\,\textsc{iv}}

\newcommand{\oiv}{[O\,\textsc{iv}]}
\newcommand{\ciii}{C\,\textsc{iii}]}
\newcommand{\oiii}{[O\,\textsc{iii}]}

\newcommand{\heii}{He\,\textsc{ii}}

\newcommand{\feii}{Fe\,\textsc{ii}}
\newcommand{\al}{Al\,\textsc{iii}}

\newcommand{\bhm}{$M_{\rm BH}$}
\usepackage[colorlinks=true, linkcolor=blue,anchorcolor=black,citecolor=blue,filecolor=black,menucolor=black,runcolor=black,urlcolor=black]{hyperref}
\usepackage{natbib}
\usepackage{subcaption}
\usepackage{makecell}
\usepackage{multirow}
\usepackage{booktabs}
\usepackage{xcolor}
\begin{document}

   \title{\texttt{AGNFormer} I: Reconstruction of AGN spectra using a probabilistic transformer model}

%
%
%

    \author{Benedict L. Rouse
          \inst{1}\corrauth{blrouse@uc.cl}
          \and
          Franz E. Bauer\inst{2}\email{franz.e.bauer@gmail.com}
          \and
          Tomasz Różański\inst{3}\email{Tomasz.Rozanski1@anu.edu.au}
          \and
          Guillermo Cabrera-Vives\inst{4,5,6}\email{guillecabrera@inf.udec.cl}
          \and
          Patricia Ar{\'e}valo\inst{7,8,5}\email{patricia.arevalo@uv.cl}
          \and
          Crist\'obal Rodrigo Donoso Oliva\inst{4,5}\email{cridonoso@inf.udec.cl}
          \and
          Pavlos Protopapas\inst{9}\email{protopapas@gmail.com}
          \and
          Gordon T. Richards\inst{10}\email{gtr@physics.drexel.edu}
          \and
          Paula S\'anchez-S\'aez\inst{11}\email{Paula.SanchezSaez@eso.org}
          \and
          Ezequiel Treister\inst{2}\email{etreister@academicos.uta.cl}
          \and
          Christian Wolf\inst{3,12}\email{christian.wolf@anu.edu.au}
          }

   \institute{Instituto de Astrofísica and Centro de Astroingeniería, Facultad de Física, Pontificia Universidad Católica de Chile, Campus San Joaquín, Av. Vicuña Mackenna 4860, Macul Santiago, Chile, 7820436
        \and
            Instituto de Alta Investigaci{\'{o}}n, Universidad de Tarapac{\'{a}}, Casilla 7D, Arica, Chile
         \and
            Research School of Astronomy \& Astrophysics,
            The Australian National University,
            Cotter Rd., Weston, ACT 2611, Australia
       \and
            Department of Computer Science, Universidad de Concepci\'on, Edmundo Larenas, Concepci\'on, 4070386, Chile
        \and
            Millennium Institute of Astrophysics, Nuncio Monseñor Sótero Sanz 100, Of 104, Providencia, Santiago, Chile
        \and
            Millenium Nucleus for Galaxies (MINGAL), Concepción, Chile
        \and
            Instituto de F\'isica y Astronom\'ia, Facultad de Ciencias, Universidad de Valpara\'iso, Gran Breta\~na 1111, Valpara\'iso, Chile
        \and
            Millennium Nucleus on Transversal Research and Technology to Explore Supermassive Black Holes (TITANS)
        \and
            John A. Paulson School of Engineering and Applied Sciences, Harvard University, Cambridge, MA, USA
        \and
            Department of Physics, Drexel University, Philadelphia, PA 19104, USA
        \and
            European Southern Observatory, Karl-Schwarzschild-Strasse 2, 85748 Garching bei München, Germany
        \and
            Centre for Gravitational Astrophysics (CGA), Australian National University, Building 38 Science Road, Acton ACT 2601, Australia
}

   \date{Received XXXX, XXXX}


  \abstract
   {}
   {We explore how an uncertainty-aware transformer-based architecture can leverage information embedded across the entire observed optical spectra of active galactic nuclei (AGN), focusing on the algorithm's ability to predict unseen or masked parts of luminous AGN spectra. This provides a direct probe of the learnable correlations between AGN continua and broad lines.
   }
   {We introduce \texttt{AGNFormer}, a transformer model trained to predict the mean expected flux and variance in masked spectral regions (major broad lines to ${\pm}10^{4}$\,km\,s$^{-1}$; missing halves), inputting rest-frame spectral fluxes and uncertainties across the entire redshift range of the SDSS DR16 Quasar Catalogue. We evaluate the performance of the model on both full (no S/N limit) and high-quality (S/N $>$ 10) spectral samples using the negative-log likelihood,
   and via comparisons with existing \civ\ and \lyman\ reconstruction algorithms.
   }
   {The model successfully reconstructs unseen AGN broad lines to better than 10-16\% (4-8\%) of the flux for the full (S/N > 10) test sets, up to an error floor of $\approx$2-6\% of the flux at S/N $\approx$ 40, while predictions for larger unseen halves grow to 12-25\% (5-15\%) of the flux the further away they are from the cut-off wavelength of the seen input spectrum.
   Predictions faithfully reproduce the broad AGN spectral diversity across the entire optical and UV quasi-stellar object (QSO) main sequence parameter spaces, including both Gaussian and Lorentzian profile regimes, \feii\ complexes, and narrow emission lines.
   Performance is similar or better compared to previous spectral reconstruction algorithms.
   }
   {The high precision of the broad-line region reconstruction demonstrates that the method successfully aggregates information across the spectrum and highlights how the AGN continuum and weaker lines/complexes have the potential to assist astronomers in the extraction of the entire wealth of information embedded in AGN spectra. \texttt{AGNFormer}'s underlying model could be a valuable tool for processing spectra from many other classes of astrophysical objects.}

   
   \authorrunning{B. L. Rouse et al.}
   \titlerunning{\texttt{AGNFormer} I: AGN spectral reconstruction}
   \maketitle
   \nolinenumbers

\section{Introduction}
Multiwavelength observations of unobscured Active Galactic Nuclei (AGN) generally present numerous tell-tale signs of accretion onto a supermassive black hole (SMBH) and subsequent reprocessing by its surrounding structure \citep[e.g.][]{alexander2012,Ramos2017}.
From these features, we have pieced together that the inner region of an AGN is composed of:
a SMBH (${\gtrsim}10^{5}$\,$M_{\odot}$);
ultraviolet (UV) to near-infrared (NIR) radiation from a geometrically thin, optically thick, multi-temperature accretion disk \citep[$\sim$1--$10^{5}$\,$r_{g}$; e.g.][]{shakura1973,1989sun, kelly2008};
Compton-upscattered X-rays from a corona of 'hot' electrons above the accretion disk \citep[$\sim$1--20\,$r_{g}$; e.g.][]{Haardt1994,wilkins2015,ricci2020};
a population of fast-moving clouds photo-ionised by the accretion disk \citep[$\approx$1000--10,000 km\,s$^{-1}$; $\sim$10s--100s light-days; e.g.][]{baldwin1997};
a population of dusty clouds, likely clumpy in nature, lying beyond the dust sublimation radius that reprocesses the accretion light as infrared (IR) emission \citep[$\sim$0.3--100 light-years;][]{krolik1988, tristram2007, nenkova2008};
and a population of lower-velocity clouds photo-ionised by the accretion disk \citep[$\lesssim$1000 km\,s$^{-1}$, $\sim$10s--1000s light-years; e.g.][]{heckman1980, ho1995,hoa1997,hob1997}.
%
%
In the highly accessible optical band, these structures manifest as a pseudo power-law continuum (accretion disk), broadened permitted emission lines (from high-velocity `broad line region' gas clouds; BLR), narrow permitted and forbidden emission lines (from the lower-velocity `narrow line region' gas clouds; NLR). The high-velocity gas clouds lie within the sphere of influence of the SMBH, and hence the permitted ionised emission lines from these clouds are encoded with information on the SMBH mass via Doppler broadening \citep{peterson1999}.


Several important correlations have been well-established among the optical emission features associated with the various structures outlined above, all of which are ultimately powered by the accretion disk.
For example, strong  correlations have been observed between the optical/UV continuum and line luminosities \citep[e.g.][]{mushotzky1984}, as well as clear connections between the UV slope and relative line strengths \citep[e.g.][]{baldwin1981,kauffman2003,kewley2001,kewley2006}
Related to this is the so-called Baldwin effect, an inverse correlation between the equivalent widths (EWs) of many broad lines and the continuum luminosity \citep[e.g.][]{baldwin1977, baldwin1978, dietrich2002}.
The EWs appear to have a stronger anti-correlation with the physically motivated Eddington ratio \citep[i.e., the "modified Baldwin effect";][]{baskin2004}.
Building on this, \citet{boroson1992} applied principle component analysis (PCA) to the 4300--5700\AA\ rest-frame spectra of Palomar-Green QSOs \citep{green1986} and showed that recovered fundamental eigenvectors (EV) are strongly correlated with physically relevant quantities; EV1 relates a correlation between FWHM$_{\rm H\beta}$-$F_{\rm \oiii}$-$EW_{\rm \feii}$; EV2  considers the inverse correlation between $F_{\rm \heii}$ and `optical luminosity'. EV1 is considered to be related to the Eddington ratio, the radio and X-ray qualities, and orientation, and it has been proposed as defining a QSO `main sequence' \citep[QMS; e.g.][]{sulentic2000b,shen2014}. A similar relation is seen in the UV between the \civ\ velocity blueshift and the EW of various UV broad emission lines \citep[e.g. \civ, \heii;][]{Sulentic2007}.
These blueshifts have been shown to be anti-correlated with the \civ\ and \heii\ EWs, suggesting that the strong winds which are responsible for the \civ\ offset are also diluting the emission from virialised gas \citep[e.g.;][]{weymann1981,richards2011,baskin2013,baskin2015,rankine2020, stepney2023}. The higher blueshifts are a likely tracer of higher Eddington ratios where more radiation pressure drives these accretion disk winds.
%
%
%
In this work, we explore the ability of a transformer-based machine learning (ML) method to extract and learn these underlying correlations across portions of the UV-optical AGN spectrum.

ML techniques  offer a powerful and structured methodology to harness this spectral information and ultimately gain physical insights. Early works leveraged PCA analysis to aid in the extraction of underlying correlations in optical and X-ray spectra of AGN \citep{mittaz1990,boroson1992,francis1992, vaughan2004}. Further works built on this using mean-field independent component analysis (MFICA), a method to extract principal components with physical meaning, that can be analysed by standard methods. These have proved successful in finding continua and AGN components and reconstructing AGN spectra in the presence of random absorption \citep{allen2013, rankine2020}. Variational autoencoders have been used for dimensionality reduction of spectra \citep{portillo2020} and can simplify spectral information for further ML analysis. They have also been invoked to estimate reduced/noiseless spectra \citep{scourfield2023}.


One particularly relevant advancement in ML has been in the area of natural language processing (NLP), driven by transformer models, which effectively overcome the limitations of recurrent neural networks (RNNs), such as difficulties with long-range dependencies. Introduced in \citet{transformer}, transformers utilise an attention mechanism that allows for parallelisation and improved handling of long sequences, making them highly successful in various generative AI models. While transformers have initially been used in language models, their flexibility has sparked interest in other fields, including astronomy \citep[e.g.][]{astromer2023,astroclip,2024leung,transformerpayne,cabrera2024, fortino2025}. Early applications of transformers in this domain have focused on stellar and low-redshift galaxy spectra, as well as light curve classification. This paper introduces a self-supervised application of transformers to AGN spectra datasets, exploring new ways to extract insights from these data using advanced deep learning techniques.

We choose to train our model via representation learning, where the model learns a representation of the data in order to extract meaningful information, through self-supervision. This method has shown strong promise in its various carnations, such as masked autoencoding, to enhance model performance \citep{lewis2019, he2021}, by using large sets of unlabelled data. Self-supervised tasks are defined by their ability to perform supervised training on a dataset that does not have typical labels; i.e., the use of an unsupervised method in a supervised manner.
For instance, \citet{fortino2025} use a self-supervised pre-training and supervised fine-tuning for the classification of supernovae from optical spectral data, whereas in this work we explore a self-supervised test of reconstructing masked regions of AGN spectra. In our case, we mask and predict the fluxes, thus treating them as labels.

A subsequent design choice was to include the flux uncertainty as an input and prediction, as deep ML models tend to be overconfident in their predictions \citep{nguyen2014,guo2017,kristiadi2020} and prior work has shown that injecting data uncertainty into the model is useful to mitigate such overconfidence \citep{Lakshminarayanan2016,kong2020}. Furthermore, introducing uncertainty into pre-trained transformer models has yielded performance gains for language tasks \citep{wu2023,chun2024} and real world tasks alike \citep{shen2025,leyton2025, rauba2026}. The inclusion of uncertainties has also been shown to enhance downstream regressions tasks during fine-tuning  \citep{li2021, oreshkin2025}.
Here, we inject and model explicitly the uncertainty on the spectral fluxes, which improves the models understanding of the difference between noisy spectra and unusual spectral features. These representations will likely be useful for any feature downstream tasks involving uncertainty or multiple epoch spectroscopy.


Here we present \texttt{AGNFormer}, in order to demonstrate the power of a transformer-based model to accurately reproduce missing spectral features or regions and leverage information that an expert would likely discard. As astronomers, we are trained to focus on certain information in a spectrum: for example, using broad lines to identify an AGN and estimate the mass of the central BH, or comparing flux ratios of \feii\ and \oiii\ to infer Eddington ratio and orientation. By contrast, an ML model does not possess this domain knowledge. In this work we highlight how critical information related to several key AGN emission lines (H-$\alpha$, H-$\beta$, \oiii, \mg, \civ, Ly-$\alpha$) is also imprinted in weaker lines and continuum components, and can ultimately be harnessed.
The outline of the paper is as follows.
Sec.~\ref{sec:method} provides a detailed description of the transformer model architecture that we deployed, while Sec.~\ref{sec:model_assessment} describes the model assessment. Sec.~\ref{sec:data} introduces the data used. In Sec.~\ref{sec:results}, we present a few key results, while several discussion topics are addressed in Sec.~\ref{sec:discussion}. Finally, in Sec.~\ref{sec:conclusions} we summarise our main results and discuss future prospects.

\section{Methodology}
\label{sec:method}

\subsection{Transformer neural network}
\label{sec:TransNN}
For this work, we use the transformer ML model, first described in \citet{transformer}. A transformer is a neural network architecture originally developed for sequence modelling, such as natural language. Its fundamental component is the self-attention mechanism, which enables each element of a sequence to attend to all others, thereby capturing both local and long-range dependencies efficiently. By stacking multiple layers of self-attention and feed-forward networks, the model learns contextualised representations of the sequence.

In our case, the sequence is not text but an AGN spectrum, represented as an ordered series of flux values across wavelength bins. We train the model with a masked autoencoding (MAE) objective \citep{he2021}, in which broad spectral regions are masked, and the transformer is tasked with reconstructing them from their surrounding context. This masking encourages the model to learn both continuum trends and the structure of spectral features. An overview of our model is presented in Figure \ref{fig:arch_plot}, where the main differences between our architecture and that of \citet{transformer} is the replacement of the classification objective with the regression of masked fluxes and uncertainties; randomly initialised register tokens and the use of solely the encoder and in that, the use of pre-layer normalisation as opposed to their use of post-layer normalisation.

We briefly describe the components most crucial to the transformer architecture, as well as those that we have modified from the original to create \texttt{AGNFormer}.


\subsubsection{Flux \&\ uncertainty embedding}
\label{sec:embedding}
Tokenisers and input embedding layers are essential aspects of transformers in an NLP context, allowing non-numeric inputs such as words to be mapped into machine-readable, learnable representations. Tokenisers are unnecessary for spectra because each spectral sample is already a numeric value. Instead, we project each scalar flux $F_\lambda\in\mathbb{R}$ and flux uncertainty $e_\lambda\in\mathbb{R}$ at wavelength $\lambda$ into a $d_{\rm model}$-dimensional embedding vector via a shared linear projection followed by a rectified linear unit (ReLU), where $d_{\rm model}$ is the dimension of the transformer:
\begin{equation}
    \mathbf{x}_\lambda = \mathrm{ReLU}(\mathbf{W} [\bar{f}_\lambda,\;e_\lambda]^{\top} + \mathbf{b}),   \qquad  \mathbf{W} \in\mathbb{R}^{d_{\rm model} \times 2}
\end{equation}
where $\mathbf{W}$ and $\mathbf{b}$ are learnable parameters (weight and bias, respectively) and $\bar{f}_\lambda$ denotes the flux perturbed by the observational uncertainty during training (Sect. \ref{sec:model_assessment}); at inference $\bar{f}_\lambda = f_\lambda$.
If a flux is masked, as described in Sect. \ref{sec:masking}, $f_{\lambda} = 0$, $e_{\lambda} = 0$ and $\mathbf{x}_{\lambda} = \mathrm{ReLU}(\mathbf{b})$.

\subsubsection{Wavelength-based positional encoder}
\label{sec:wavelength_encoding}
To enable the transformer to interpret correlations across a spectrum in a physically meaningful way, we employ a wavelength-based positional encoder. Embedding wavelength information in conjunction with its associated flux allows the model to reason about the relative separation of features in wavelength space, rather than assuming a uniform spacing. This is particularly important for astronomical spectra, where wavelength grids may vary across instruments or redshift ranges.

In the original transformer architecture \citep{transformer}, positional encoding information is injected through a fixed sinusoidal encoding applied to each token index $p$:
\begin{align}
    \text{PE}(p, 2i) &= \sin\!\left(p / 10000^{2i/d_{\text{model}}}\right), \label{eq:posenc1} \\
    \text{PE}(p, 2i+1) &= \cos\!\left(p / 10000^{2i/d_{\text{model}}}\right),
\end{align}
where $d_{\text{model}}$ is the embedding dimension and the exponent the range of frequencies. This encoder receives $p = 0, 1, 2, ..., L-1$ where $L$ is the length of the input sequence. Here, we substitute this token index for a continuous, physical wavelength, $log_{10}(\lambda)$. This requires the sinusoidal encoding to be defined as
\begin{align}
    \text{PE}(\lambda, 2i) &= \sin\!\big[(\lambda - \lambda_{\text{min}})\,\omega_i\big],\label{eq:posenc2} \\
    \text{PE}(\lambda, 2i+1) &= \cos\!\big[(\lambda - \lambda_{\text{min}})\,\omega_i\big],
\end{align}
where the angular frequencies $\omega_i$ are sampled geometrically between
\begin{align}
    \omega_{\text{max}} &= 1 / \Delta\lambda, \\
    \omega_{\text{min}} &= \omega_{\text{max}} \times 10^{-4},
\end{align}
and $\Delta\lambda$ is the average wavelength spacing of the spectrum. This range spans four orders of magnitude in frequency, ensuring sensitivity to both short and long range variations in wavelength space.

If the wavelength sampling is uniform and integer-valued, this encoding reduces exactly to that of \citet{transformer}\footnote{A mathematical proof can be found in Sec.~\ref{app:proof}.}. However, defining the positional encoding in physical units allows the model to naturally accommodate for the sources being shifted to the rest-frame and therefore on different wavelength grids. It also allows for irregular wavelength sampling or logarithmic wavelength grids without the need for resampling. This wavelength encoding will allow our model to be more versatile when receiving data across large redshift ranges and from distinct spectrographs with varying spectral resolutions. Furthermore, this positional encoder couples each flux and wavelength value together, meaning that the entire spectrum can be shuffled or even entirely reversed, as would be the case if working in frequency space over wavelength space, and the same results would be produced.
\begin{figure*}
    \includegraphics[width=\textwidth, trim = 0 0 0 0, clip]{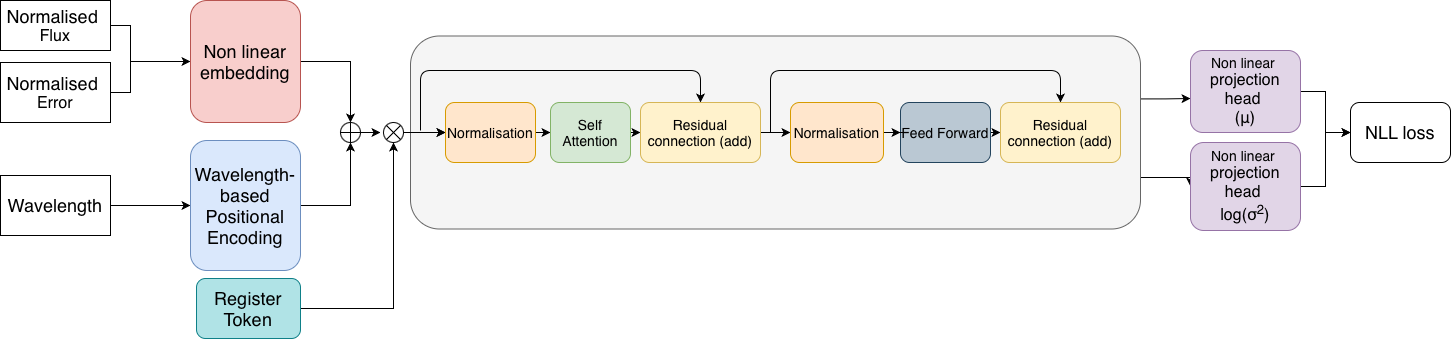}
    \caption{Architecture of the probabilistic transformer used in this work. We apply a shared non-linear embedding to the fluxes and their uncertainty. The wavelengths are introduced via the wavelength-based positional encoder. The encoder is modified from the original presented in \citet{transformer}, to perform layer normalisation before entering the Sublayer (attention or feedforward). There are two non-linear projection heads that allow for the calculation of the mean flux ($\mu$) and the log variance [log($\sigma^{2}$)]. Both of these are fed into the NLL loss.}
    \label{fig:arch_plot}
\end{figure*}
\subsubsection{Attention mechanism}
\label{sec:attention}
A fundamental component of the transformer is the dot-product attention mechanism, which builds upon work on soft alignment in sequence models \citep[e.g.][]{bahdanau2014}, as well as early computer vision attention work \citep[e.g.][]{attention}. These prior works aimed to provide human-interpretable outputs in the network phase of the algorithm, aiming to avoid the 'black box' nature of deep neural networks. \citet{transformer} later applied the attention mechanism to LMs.

In the transformer, each token in the sequence is first projected into three distinct vectors: a query (Q), a key (K), and a value (V). The query represents what the token is “asking for,” the key represents what information the token “offers,” and the value is the actual content that can be passed along. In self-attention, the model compares the query of each token to the keys of all tokens in the sequence, producing attention scores that are normalised into probabilities. These probabilities determine how strongly each token’s representation should incorporate information from the values of other tokens. 

The attention probabilities represent, for each position in the spectrum, how much influence every other position has in informing its updated representation. 
The final attention output is computed via:
\begin{equation}
    \rm Attention(Q,K,V) = \rm softmax\Bigg(\frac{QK^T}{\sqrt{d_{k}}}\Bigg)V,
\end{equation}
where softmax($x$) is:
\begin{equation}
\mathrm{Softmax}(\mathbf{x}_i) = \frac{e^{x_i}}{\sum_{j=1}^{K} e^{x_j}}
\end{equation}
so that the values serve as the routed content, directed by the relationships between Q and K. In practice, we employ multi-head attention, which involves performing this interaction many times in parallel and subsequently concatenating the results. For a comprehensive look into multi-head attention, we refer the reader to \citet{transformer}. 

In the context of using a transformer to predict masked fluxes of a spectrum, this example may allow some insight on Q, K and V: Q asks what information a masked spectral feature, e.g. \hb\, needs in order to be predicted. K, corresponding to a spectral feature from unmasked wavelengths (for example continuum regions or nearby narrow lines) encodes what information can be offered that aligns with the masked feature. The dot product of Q and K determines how compatible the two are; a large attention score indicates that the continuum position is compatible and informative for predicting \hb. Once the attention scores have calculated which positions are relevant, V is introduced, which contains the actual content of those continuum positions with high compatibility, in order to build the final representation of the masked flux.

\subsubsection{Encoder}
\label{sec:encoder}

Our implementation of the transformer can be seen in Fig. \ref{fig:arch_plot}. Similar to BERT \citep{devlin2018}, the model is an encoder-only transformer with small modifications as follows.
The encoder takes the input sequence and attempts to capture the significance and context of the input by mapping it onto hidden states.
This encoder comprises a user defined number of layers, which can be broken down into two sub-layers. The first sub-layer is the multi-head self-attention mechanism, described in Sect. \ref{sec:attention}.
The second sub-layer is the feedforward neural network, which receives the representation of each value from the attention block and passes them through fully connected layers and applies non-linear transformations to each value. Both of these sub-layers contain a residual connection, which allows the input to the sub-layer to be fed into the attention or feedforward and summed to the output of these mechanisms. This residual connection allows for stable gradient flow and training of deep, expressive models. 


In addition to the residual architecture, the second technique used for stabilising training is layer normalisation. Recent implementations with larger models and training sets have found that "post-normalisation" (Post-LN), following the format proposed in \citet{transformer}, produces highly unstable gradients and requires very careful hyperparameter tuning \citep{xiong2020}. Post-LN is typically of the form \hbox{$x$ $=$ LayerNorm[$x$ + Sublayer($x$)]}, where x is the input vector, LayerNorm is the layer normalisation and Sublayer is either the self attention layer or feedforward layer.  Instead, transformer encoders now typically adopt a "pre-normalisation" \citep[Pre-LN; e.g.][]{baevski2018,xiong2020} as \hbox{$x$ + Sublayer[LayerNorm($x$)]}, which leads to a smoother gradient flow and results in more stable training, especially for large models and training datasets. Our initial experiments with Pre-LN showed that it provided a 20\% reduction in training loss compared to Post-LN.

In unison, these sub-layers output a sequence which provides information on both the local and global dependencies of the input sequence.

\subsubsection{Registers}

Attention maps are not often as informative or interpretable as one might hope. For example, \citet{darcet2023} show that for image classification with vision transformers, the attention map will often highlight background pixels that contain zero information on the object being classified and thus create artefacts in the attention maps; this is exacerbated for larger models. Their proposed solution to this is to introduce what they refer to as "registers". These are essentially randomised, empty tokens that can be concatenated to the end of the input sequence, which allow the model to perform calculations that are not necessarily useful in attention visualisation. These tokens are disregarded when the attention maps are visualised, yielding attention maps that are more concentrated in relevant regions of the input image. \citet{darcet2023} find that for classification, more register tokens are better. They test a range of possibilities from 1 to 16 tokens. In Fig. \ref{fig:reg_lr}, we show the result of our hyperparameter optimisation as a function of the number of register tokens and learning rate. We find that 1 token outperforms all other options, including 0 tokens. 

Formally, we define the register token as a learnable parameter matrix: $\mathbf{R} \in \mathbb{R}^{N_\text{reg} \times d_\text{model}}$,
initialised from a normal distribution $\mathcal{N}(0, 0.01)$. At each forward pass, the register tokens are appended to the encoder input embeddings: $\mathbf{X} \in \mathbb{R}^{L \times d_\text{model}}$, yielding the augmented sequence:
$\tilde{\mathbf{X}} = [\mathbf{X};\, \mathbf{R}]$.

\subsubsection{Non-linear projection heads}
Finally, we pass the outputs of the encoder into two separate multi-layer perceptron (MLP) projection heads that consist of a linear layer, activation and a second linear layer. One projection head is dedicated to computing the mean ($\mu$) flux and the second, the log variance [log($\sigma^{2}$)] of that flux.
If we were to use a shared projection head, both $\mu$ and $\sigma^{2}$ would be computed from the same latent representation, forcing them to be linear projections of the same features.
Separating the heads allows the intermediate features to be computed independently and specialise differently, 
allowing the model to be more expressive for both outputs.

We model the predictive distribution of each masked pixel as an independent Gaussian. This is a deliberate simplification: it cannot capture multimodal outcomes (e.g. the presence or absence of broad absorption in \civ) or correlated uncertainty across a line profile. Relaxing this assumption (e.g. mixture-density or quantile outputs) is left to future work.
\subsection{Masking}
\label{sec:masking}
To perform self-supervised training, we apply a masking procedure to the input spectra and the model is trained to reconstruct masked parts of the spectra. In training, we perform a combination of random and fixed masking as incorporating random masking will ultimately create a better generalised model than if we trained on fixed masking alone. In inference, we solely perform fixed masking in order to provide a consistent sample for comparison across various runs of the model. The masking in training is split into random pixel, random chunk and fixed masking, adopting a 35:40:25 split of these methods respectively.

The random pixel masking works by randomly masking $\in [0.1,0.2]$ of the pixels in a spectrum, ensuring that pixels are never adjacent to one another. Random pixel masking teaches the model about shapes of the spectra, and it will likely just learn to interpolate.

The random chunk masking randomly draws a number of chunks of adjacent pixels (maximum of 3 chunks) and a percentage of the spectrum within a chunk $\in [0.07,0.15]$ to mask. We treat the spectrum as a continuous loop so that an individual chunk may partially mask the end and beginning of the spectrum. Random chunk masking pushes the model to leverage information outside the masks to infer the fluxes within the masked region, which should give the model a strong understanding of how features relate and influence each other across the spectrum.

The random masking is designed so that the probability of a token being masked is equal across the entire spectrum. We also require that the mask vary each time the spectrum is loaded into the model (each training epoch); providing the additional benefit that each spectrum can generate multiple unique training spectra.

For fixed masking, a set of user defined wavelength ranges are masked in each spectrum. We adopt a width of 20,000\,km\,s$^{-1}$, masking 10,000\,km\,s$^{-1}$ either side of the laboratory central wavelength of the broad line that is subsequently predicted. The fixed masking helps to ground the model in the scientific goal of leveraging information from the continuum of the spectrum to infer broad line region information.

In each masked region, flux and error values are replaced with zeroes and \texttt{AGNFormer} reconstructs these missing values. Thus, the reconstruction loss is calculated solely across these masked regions. 
We choose to apply a corresponding mask in the attention layers, whereby the masked wavelengths attention score is set to -$
\infty$. Doing so forces the model not to attend to these masked regions, thus only drawing from information in the unmasked spectrum. We also apply attention masking for zero padded regions of the spectrum, which we describe in Sect. \ref{sec:preprocessing}. 

%
\section{Loss function \& model training}
\label{sec:model_assessment}

The goal of this paper is to train a transformer model to perform masked flux/flux uncertainty prediction of keys portions of AGN/QSO spectra with the intention of using this as a pretrained model which provides the initial weights for future fine-tuning tasks. These tasks could include spectral parameter prediction or some latent representation with photometry, light curves or images. We view this work as an introduction to the model's potential, wherein we aim to train a very broad, generalised model.

We ultimately adopt a model with 4 attention heads, 8 encoder layers (1 pass through grey box of Fig.~\ref{fig:arch_plot} $=$ 1 layer), a model dimensionality of 256 and the hidden feedforward dimensionality of 1024. This selection of hyperparameters allows for enough complexity within the model to build the representation between continua and broad lines while minimising training time and over-fitting.

During training, we sample two disjoint index sets,
\begin{equation}
\qquad
\mathcal{U} \subset \{1,\ldots,N\},
\qquad
\mathcal{M} \subset \{1,\ldots,N\}.
\qquad
\end{equation}
where N is the sequence length, $\mathcal{U}$ contains the unmasked flux as input to the attention mechanism, which learns the relationships between spectral features and infers the missing masked regions, $\mathcal{M}$, sampled as described in Sect.~\ref{sec:masking}. We also include the uncertainty on the unmasked flux values in order for the model to have a picture of the aleatoric error.
The model receives the context set
\begin{equation}
X_{\mathcal{U}}
=
\left\{
\left(\lambda_j, f_j, e_j \right)
\right\}_{j\,\in\,\mathcal{U}},
\end{equation}
where $\lambda_{j}$ is the wavelength of pixel j, $f_{j}$ is the observed flux and $e_{j}$ is the known observational uncertainty, together with the target wavelengths
\begin{equation}
\Lambda_{\mathcal{M}}
=
\left\{
\lambda_i
\right\}_{i\,\in\,\mathcal{M}} .
\end{equation}

For each target pixel $i\,\in\,\mathcal{M}$, the neural network predicts a mean flux and an intrinsic predictive uncertainty,
\begin{equation}
\left(
\mu_{\theta},
\sigma_{\theta}
\right)_{i}
=
\rm \texttt{AGNFormer}_{\theta}
\left(
X_{\mathcal{U}},
\Lambda_{\mathcal{M}}
\right)_i .
\end{equation}
Here $\theta$ represents the trainable parameters of \texttt{AGNFormer}; $\mu_{\theta}$ is the predicted mean flux while $\sigma_{\theta}$ describes additional model uncertainty not contained in $e_{j}$.

In order to make the model explicitly noise aware, we perturb the input fluxes with Gaussian noise scaled by the reported uncertainties,
\begin{equation}
    \bar{f_{j}} = f_{j} + \epsilon_{j} \qquad \epsilon_{j}\sim\mathcal{N}(0, e^{2}_{j}),
\end{equation}
where the flux uncertainty is normalised by the same factor as the flux (see Sect. \ref{sec:preprocessing}). The noise is resampled at every epoch, so the model never sees the exact same realisation of a spectrum twice. This perturbation encourages the spectrum to learn the underlying spectral distribution and discourages it from treating any flux value as exact. Both the perturbed flux and the uncertainty are provided as input (Sect.~\ref{sec:embedding}), allowing the model to distinguish an unusual feature measured with low uncertainty from one that is consistent with noise. Masked pixels (Sect.~\ref{sec:masking}) are excluded from the perturbation as both flux and uncertainty are set to zero.

The model is then trained by minimising the heteroscedastic Gaussian negative log-likelihood (NLL), defined as,
\begin{equation}
\label{eq:loss_compact}
\mathcal{L}_{\mathrm{NLL}}
= -\frac{1}{|\mathcal{M}|} \sum_{i\,\in\,\mathcal{M}}
\log \mathcal{N}\!\left(f_i \;\middle|\; \mu_{\theta,i},\;
\sigma_{\theta,i}^{2} + e_{i}^{2}\right),
\end{equation}
where $\mu_{\theta,i}$ and $\sigma_{\theta,i}^{2}$ are the mean and variance
predicted by the model for pixel $i$, and $\mathcal{M}$ is the set of masked
pixels. Expanding,
\begin{equation}
\label{eq:loss}
\mathcal{L}_{\mathrm{NLL}}
= \frac{1}{|\mathcal{M}|} \sum_{i\,\in\,\mathcal{M}}
\left[
\frac{\left( f_{i} - \mu_{\theta,i} \right)^{2}}
     {\sigma_{\theta,i}^{2} + e_{i}^{2}}
+ \log\!\left(\sigma_{\theta,i}^{2} + e_{i}^{2}\right)
\right],
\end{equation}
which reduces to the mean squared error (MSE) if $\sigma_{\theta, i}$ is not predicted and $e_{i}$ is 1, as is the case for an independent normal Gaussian noise. Therefore, the first term in the NLL loss is the MSE weighted by inverse variance. NLL allows a distribution to be modelled, as opposed to the fixed-variance case with MSE, meaning that in low-noise cases, an incorrect prediction is penalised more than poor predictions in high-noise regions. This treatment of the flux uncertainty results in a model that focuses more capacity into predicting in regions where the flux has a higher certainty rather than expending energy in predicting in noisy regions, which has led to a substantial improvement in the models predictive power over a comparative deterministic model.

We train the model over 1,000,000 steps, using an AdamW optimiser \citep{loshchilov2017} and a warm-up-stable-decay (WSD) scheduler, which we discuss in detail in Appendix~\ref{app:scheduler}.
In ML tasks, an epoch is defined as one pass through the entire training dataset, whereas steps refer to the number of batches processed in the training set. In our case, the batch size is 16 and therefore 16 sources are analysed in order for the weights to be updated.
We apply a 70:27:3 training:validation:testing split to the data. In testing, we apply the fully trained models to unseen spectra with masked regions, in order to reconstruct fluxes. We then compare the predicted fluxes to the observed values.

\section{Data}
\label{sec:data}
Transformers typically require large ($\gtrsim$$10^{5}$ sources) training sets to successfully learn relevant patterns in the data. Here, we start with the catalogue of spectra and quasar properties from SDSS DR16 \citep[DR16Q,][]{wu2022}, the largest spectroscopically confirmed QSO catalogue available at the start of the project. We also test on a small set of spectra of high-redshift QSOs from VLT X-Shooter \citep{vernet2011}.

\begin{figure}
    \hspace{-0.15cm}
    \includegraphics[width=1.02\linewidth, trim=5 0 30 35, clip]{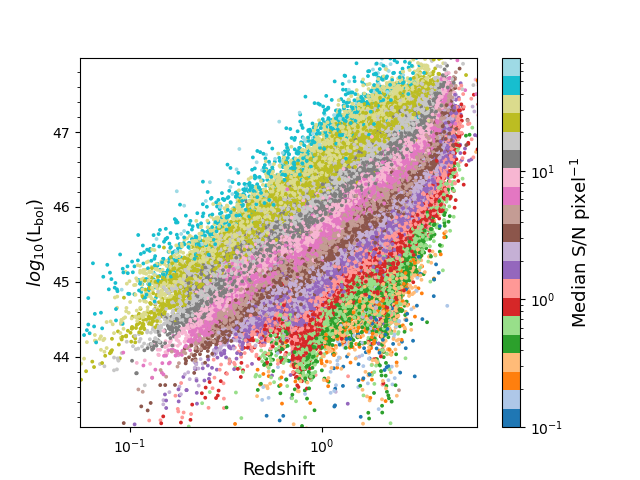}
    \caption{Distribution of redshift vs bolometric luminosity, coloured by the median S/N per pixel of the raw SDSS spectra for the entire SDSS DR16Q catalogue of \citet{wu2022}, highlighting the broad parameter space of model training.}
    \label{fig:dr16z}
\end{figure}

DR16Q provides fitted spectra for 750,414 broad-line quasars, over a redshift range of $0.1 < z < 7$, as displayed in Fig.~\ref{fig:dr16z}. 
For the present work, we perform a luminosity selection which involves removing all sources below $log_{10}(L_{5100}) = 43.5$, in a simple attempt to remove sources with dominating host galaxies. In a future work we plan to test the capability of the model to extract the host galaxy spectrum from the QSO spectrum. In an attempt to remove broad absorption line (BAL) QSOs from training, we exclude any source with $P_{BAL} > 0.0$, as determined by \citet{lyke2020}\footnote{$P_{\rm BAL}$ is tuned to high-ionization (HiBAL) troughs from \civ+\siiv; classifications become less robust for low-ionization (Lo)BALs associated with \mg, \al, and \feii. The catalog provides per-object discrete values of -1 (\civ+\siiv\ trough diagnostic region falls in the BOSS/eBOSS coverage, i.e. 1.57 $\lesssim$ z $\lesssim$ 5.6), 0 (non-BAL), 0.5 (weak absorption present with $\lesssim$\,2000\,km\,s$^{-1}$; 'mini-BAL'), 0.75 (weak absorption present with $>$\,2000\,km\,s$^{-1}$); 'mini-BAL'), 0.9 (strong absorption present, but with $\lesssim$\,2000\,km\,s$^{-1}$), 0.95 and 1 (strong absorption present with $>$\,2000\,km\,s$^{-1}$).}. We explore the reconstruction of these sources in Sects. \ref{sec:bl_results} \& \ref{sec:comparingtoICA}.

\begin{figure}

      \includegraphics[width=1.0\linewidth, trim=20 11 10 15, clip]{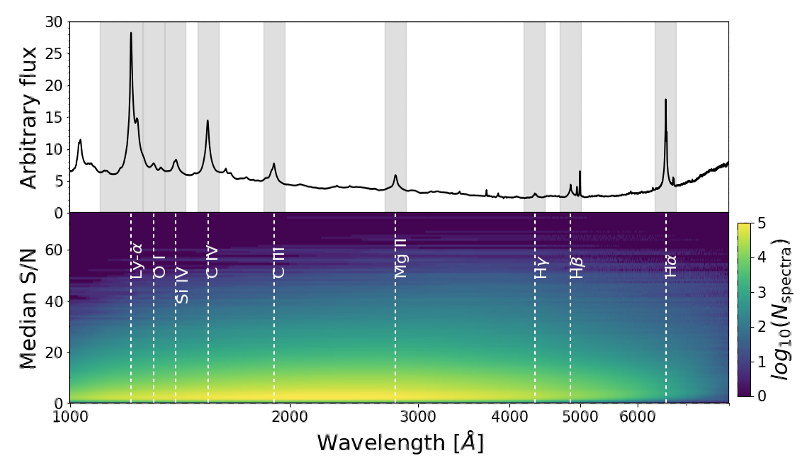}

    \caption{\textit{Top}: Median spectrum (per wavelength bin) from the entire SDSS DR16Q, with the broad lines analysed in this work highlighted in grey. \textit{Bottom}: Distribution of the rest-frame wavelength coverage vs median S/N of the spectrum. Each of the vertical lines denote interesting broad lines that we mask in testing. Sources with lower S/N ($<$10) and within the rest-frame wavelength range of 1500--4000$\AA$ dominate amongst the full sample.}
    \label{fig:sn_vs_wav}
\end{figure}

We do not make any additional signal-to-noise ratio (S/N) cuts, allowing the transformer to be trained on the entire catalogue, but acknowledge that it may ultimately be beneficial to train on only higher S/N spectra. In the bottom panel of Fig. \ref{fig:sn_vs_wav}, we plot the distribution of sources across rest-frame wavelength and S/N. It is clear that there is a heavy skew of sources populating 0 $<$ S/N $<$ 10 and that the \ha\ line is observed far less frequently than the remaining bluer eight highlighted broad lines.




We perform a test with spectroscopy from X-Shooter, a near-UV/optical/NIR spectrograph at the ESO Very Large Telescope (VLT). It has an achievable resolution between R$\approx$3200--18400 (dependent on slit size and wavelength) and spectral coverage from 0.3--2.5$\mu m$. \citet{greig2024} examined a random sample of 30 spectra from the high-quality XQ-100 sample \citep{lopez2016}, in a redshift range 3.5 $\leq$ z $\leq$ 4.5, for a \lyman\ reconstruction challenge. In Sec. \ref{sec:lyman_reconstruction}, we test on the same 30 sources to directly compare the output of our model with the \lyman\ reconstruction algorithms presented in \citet{greig2024} \footnote{They also randomly select 30 BOSS spectra from the same redshift range which we also compare to.}.

\subsection{Pre-processing}
\label{sec:preprocessing}
We do not perform any spectral resampling, but input the raw rest-frame spectrum straight into our model, only restricting each spectrum to be 4500 pixels long. Some spectra may be slightly longer than this and are therefore cut off at the red end and some may be slightly shorter, thus we apply zero padding to ensure they have a sequence length of 4500, due to the requirement of input sequences in a transformer being equal length.

A crucial step in ML analysis is normalising input data before training. For our specific task, prediction of masked spectral data, the choice of normalisation must avoid information leakage and be applicable uniformly across the dataset. For example, applying z-score normalisation to the entire spectrum can leak information when predicting masked regions, since the normalisation incorporates those masked values. To prevent this, we adopt a modified Euclidean normalisation:
\begin{equation}
    x = f\ \bigg/ \sqrt{\frac{1}{|\mathcal{U}|} \sum_{j\,\in\,\mathcal{U}} f_{j}^2}, \
\end{equation}
where $x$ is the normalised flux, $f$ the original flux, $|\mathcal{U}|$ is the number of unmasked pixels and the root-mean-square (RMS) factor is computed only over unmasked regions. Including masked values, e.g. from broad lines or continua, in the RMS factor would leak information into the model. Our approach preserves the benefits of Euclidean normalisation, such as reduced luminosity variance and retention of spectral features, whilst ensuring masked regions do not influence the normalisation. The flux uncertainty is also normalised using the same RMS factor, computed from the fluxes.
We shift all spectra to the rest-frame, using the quoted redshift from SDSS DR16Q \citep{wu2022}, where $\approx$90\% come directly from the SDSS pipeline \citep{bolton2012} while the remainder have a velocity shifted re-estimation, using the method described in \citet{shen2016}, which corrects for the average velocity offset.

\subsection{QSO MS}
\label{sec:data_qsoms}
\begin{figure}

  \begin{subfigure}[b]{0.53\textwidth}
  \hspace{-0.35cm}
     \includegraphics[width=1.1\textwidth, trim=15 10 0 65, clip]{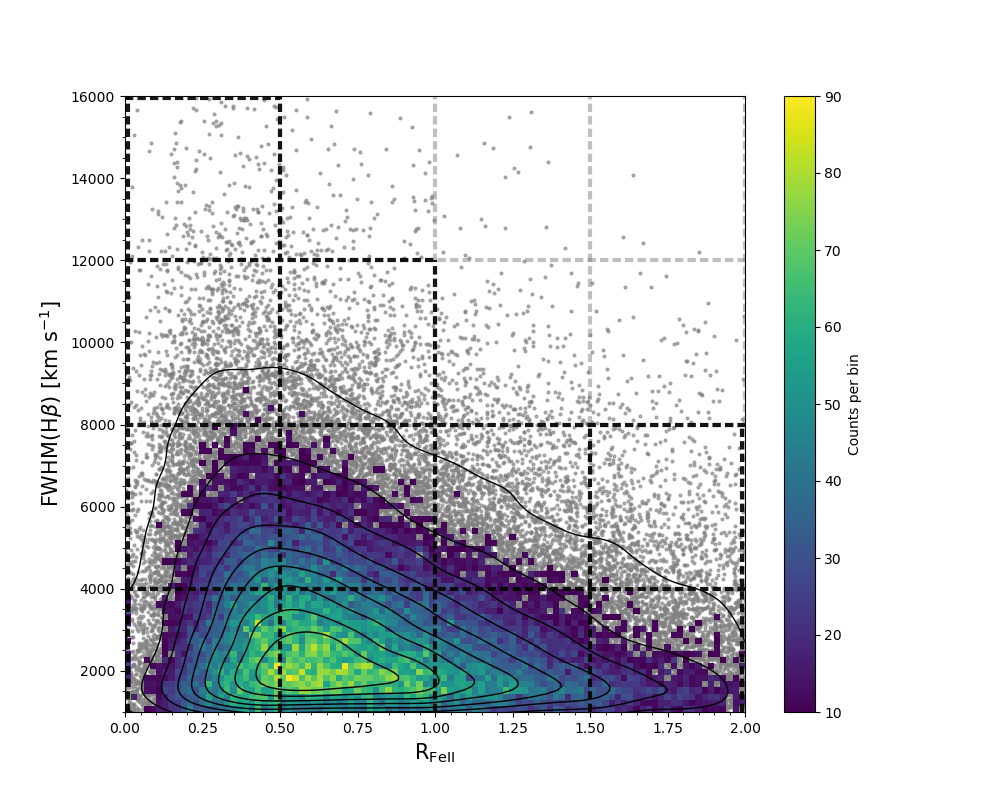}
  \end{subfigure}

  \begin{subfigure}[b]{0.53\textwidth}
  \hspace{-0.35cm}
      \includegraphics[width=1.1\textwidth, trim=15 25 0 65, clip]{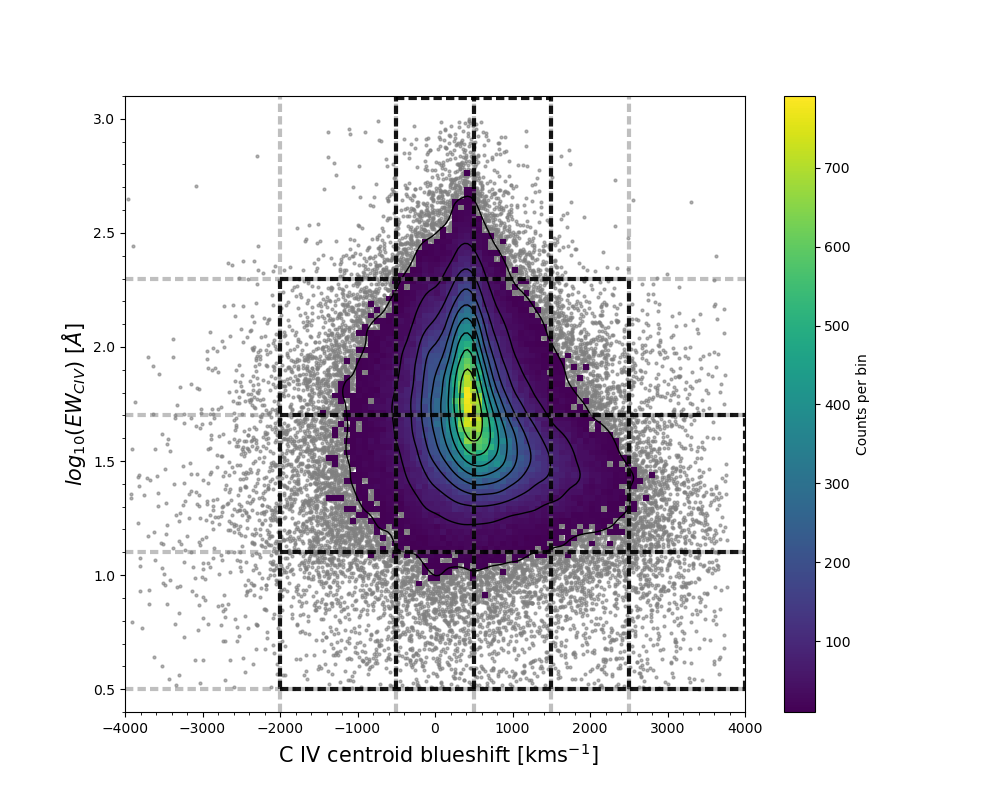}

  \end{subfigure}
    \caption{\textit{Top}: Optical-QSO Main Sequence (QMS; also known as EV1) showing $\rm R_{\feii}$ ($\rm EW_{\feii}$/$\rm EW_{H\beta}$) vs. $\rm FWHM_{H\beta}$, adopting the same grid spacing as defined by \citet{sulentic2002} to split the population. The line profiles of the sources that make up this plot can broadly be modeled by a Lorenztian profile when FWHM $<$ 8000 km$s^{-1}$ and a Gaussian when FWHM $>$ 8000 km$s^{-1}$.  \textit{Bottom}: UV-QMS showing \civ\ velocity offset ($\rm v_{\civ}$) vs. EW [$\log{\rm (EW_{\civ})}$]. Black dashed lines outline grids analysed in Sec. \ref{sec:res_qsoms} to confirm that \texttt{AGNFormer} accurately reconstructs features across these important phase spaces; sparsely populated bins are currently ignored to reduce complications with low-number statistics. Grey dots denote individual sources, in regions too sparse to be covered by the contour plot.}
    \label{fig:density_plots}
\end{figure}
\begin{figure*}
    \begin{subfigure}[b]{0.33\textwidth}
        \includegraphics[width=\linewidth, trim = 20 20 40 50, clip]{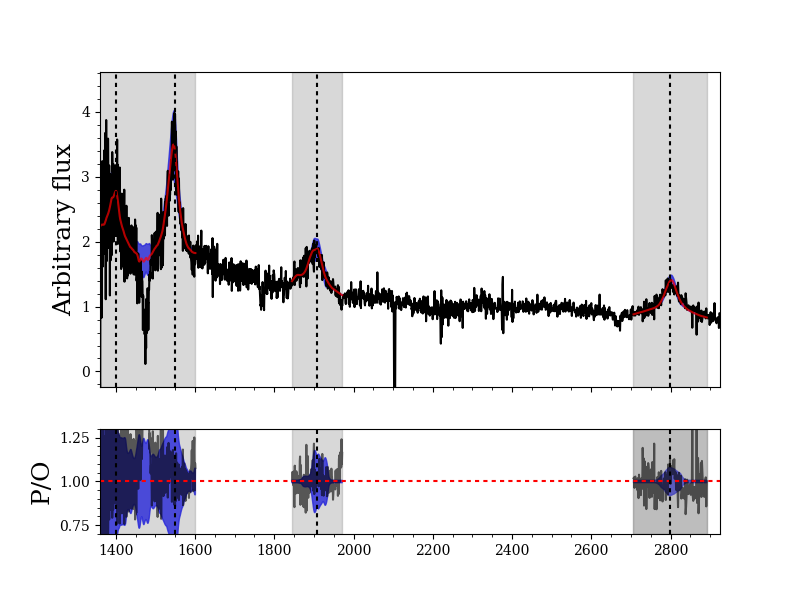}
    \end{subfigure}
    \begin{subfigure}[b]{0.33\textwidth}
        \centering
        \includegraphics[width=\linewidth, trim = 20 20 40 50, clip]{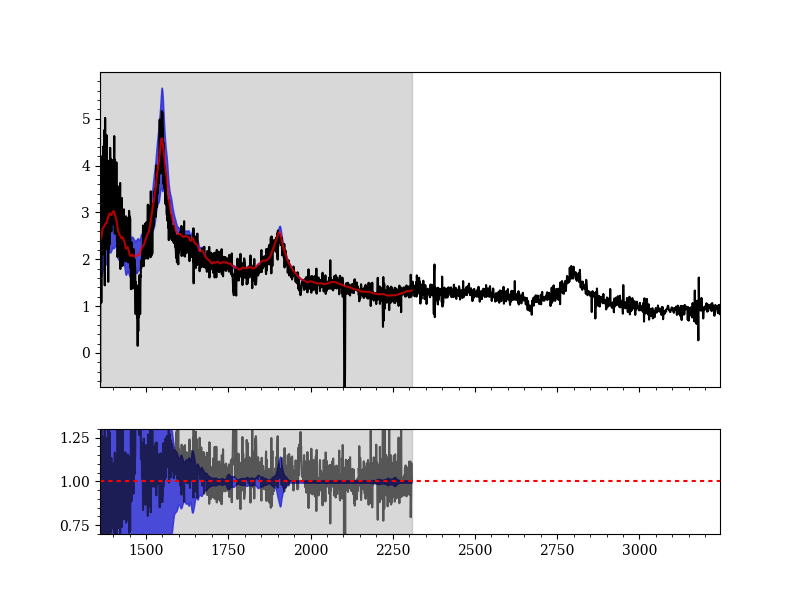}
    \end{subfigure}
    \begin{subfigure}[b]{0.33\textwidth}

        \includegraphics[width=\linewidth,trim = 20 20 40 50, clip]{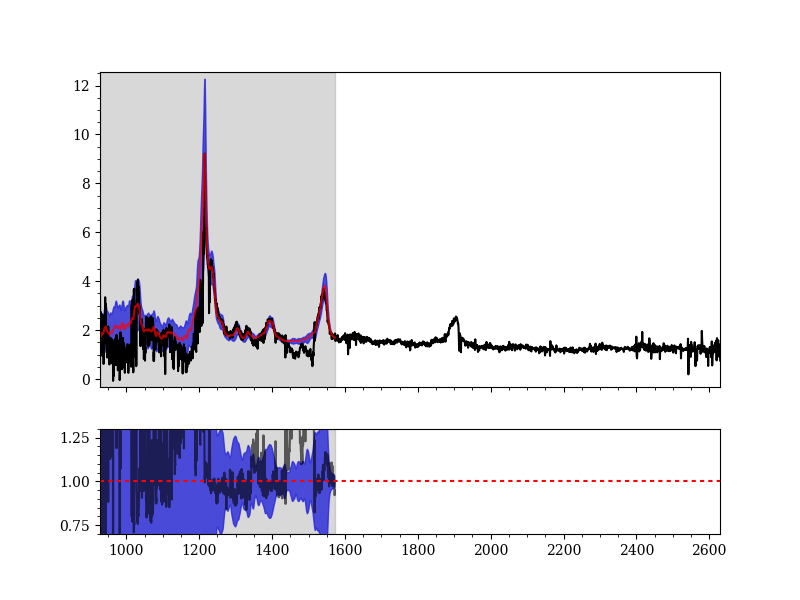}
    \end{subfigure}
    \begin{subfigure}[b]{0.33\textwidth}
        \includegraphics[width=\linewidth, trim = 20 0 40 45, clip]{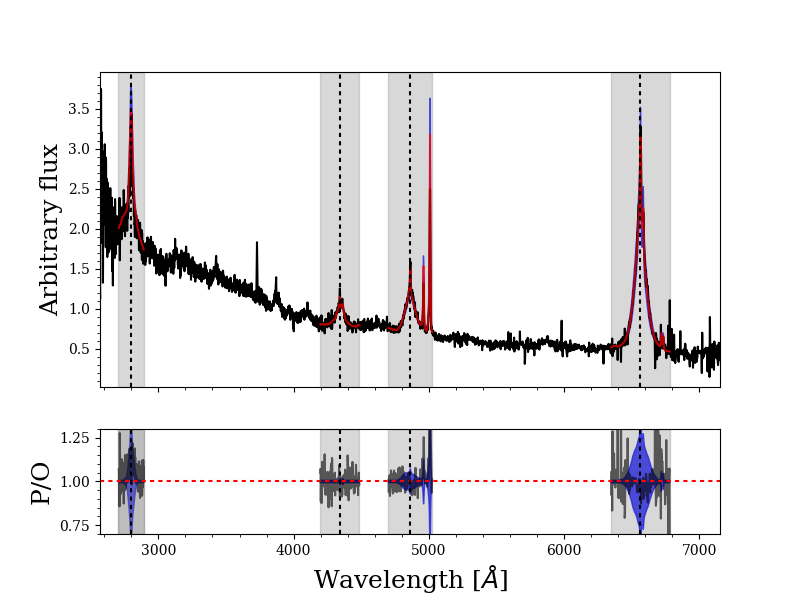}
    \end{subfigure}
    \begin{subfigure}[b]{0.33\textwidth}
        \centering
        \includegraphics[width=\linewidth, trim = 20 0 40 45, clip]{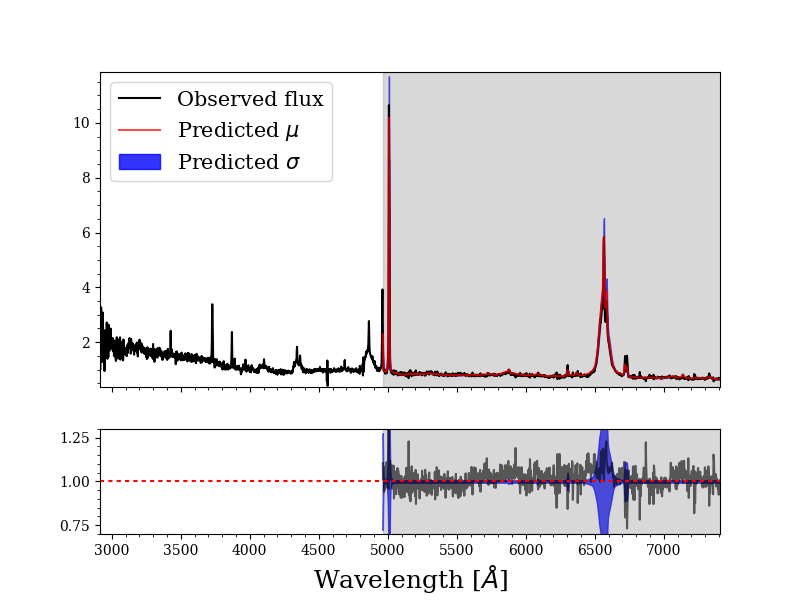}
    \end{subfigure}
    \begin{subfigure}[b]{0.33\textwidth}
        \centering
        \includegraphics[width=\linewidth, trim = 20 0 40 45, clip]{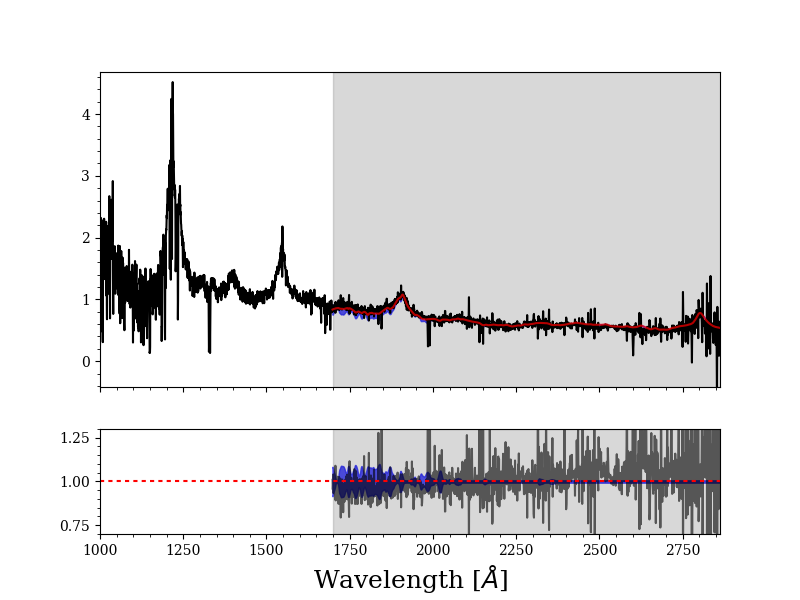}
    \end{subfigure}
    \caption{Example DR16Q spectra (observed in black) and spectral reconstructions (mean prediction $\mu$ in red, 1$\sigma$ prediction in blue), either for broad line windows (left spectra; centres of broad lines are marked by vertical dashed lines while  masked $\pm$10,000\,km\,s$^{-1}$ ranges around centres shown in grey) or spectral halves (middle and right spectra; masked and predicted portions denoted by grey regions) denoting broad lines.
    For each spectrum, \textit{top panels} show arbitrary fluxes, while \textit{bottom panels} show predicted/observed flux ratios (grey) and uncertainties (blue).
    \siiv\ and \civ\ are masked together in one large mask that encompasses the range from 1300--1600\AA\ in order to account for all BAL absorption ranges (discussed further in Sec. \ref{sec:comparingtoICA}).}
    \label{fig:singlespec}
\end{figure*}
Here we highlight two important QMS amongst the dataset that we will refer to later. Our goal is to ensure that the model learns informative representations across the entire MS, despite the inherent data imbalance that can be seen in Fig. \ref{fig:density_plots}.

The first QMS is created by separating the QSOs into distinct subsets across EV1, based on \hb\ FWHM and the ratio of \feii\ to \hb\ EW, as performed in \citet{sulentic2002}.
For low FWHM ($<$ 5000\,km\,s$^{-1}$) lines, the broad line profile is Lorentzian, but with increasing \hb\ FWHM, the broad line can be better modelled by a Gaussian \citep{kollatschny2013}. We want to test that the model is capable of predicting both profiles correctly. Unfortunately, as can be seen in Fig. \ref{fig:density_plots}, the number of sources with Lorentizan profiles dominate over those with Gaussian profiles, and thus represent a strong test of the \texttt{AGNFormer}'s ability to extrapolate its learning to low density parameter space.

A similar QMS exists in the UV, between the velocity offset ($v_{\rm \civ}$) and EW of \civ, which together trace the relative contribution of disc-driven winds to the emergent line profile, where large blueshifts ($v_{\rm \civ} \gtrsim +1000$\,km\,s$^{-1}$) and reduced EWs ($\log{\rm (EW_{\civ})}\lesssim 1.7$) are associated with stronger outflows and higher Eddington ratios \citep[e.g.,][]{Sulentic2007}. As such, it is an important parameter space in the UV for the model to learn over. In Fig. \ref{fig:density_plots} (bottom panel), we show the density of points in this parameter space, where a clear concentration of sources around $v_{\rm \civ} \sim +400$\,km\,s$^{-1}$ and $\log{\rm (EW_{\civ})} \sim 1.7$ can be seen.


\section{Results}
\label{sec:results}
We trained \texttt{AGNFormer} over the entirety of SDSS DR16Q and here perform a few specific tests, to highlight its ability to predict and reconstruct masked regions of spectra using often discarded portions of AGN spectra such as the continuum, weaker emission lines, and/or the blended Fe {\sc ii} and Balmer line complexes. One test involved supplying an input spectrum with all major emission lines (broad \oi, \siiv\ + \oiv, \civ, \ciii, \mg, \hg, \hb\ and \ha\ and narrow \oiii\ 4960\AA\ + 5008\AA) masked, and predicting them based on the remainder of the supplied spectrum (Sect.~\ref{sec:bl_results}).
Another test involved masking half of the input spectrum and using the unmasked half to predict the unseen half (Sect~\ref{sec:half_spec_res}). Finally, we wanted to ensure that the model accurately predicts spectra across the broad parameter space of the QSO main sequence (Sect.~\ref{sec:res_qsoms}). 

\subsection{Broad line estimation}
\label{sec:bl_results}

\begin{table}

    \begin{tabular}{l|r|r|c|c}
        \makecell{Broad \\line} & \makecell{Source \# \\ (full)} & \makecell{Source \# \\ (S/N > 10)} & \makecell{median S/N \\(full)} & \makecell{median S/N  \\(S/N > 10)} \\
        \hline
        \oi & 304047 & 31096 & 3.41 & 13.37\\
        \siiv & 371269 & 39285 & 3.47 & 13.25 \\
        \civ & 468244 & 50757 & 3.55 & 13.25 \\
        \ciii & 621871 & 72713 & 3.69 & 13.20\\
        \mg & 652784 & 88005 & 3.99 & 13.28 \\
        \hg & 238229 & 43777 & 4.74 & 13.62\\
        \hb & 153597 & 31738 & 5.11 & 13.86 \\
        \ha & 22887 & 9546 & 8.61 & 15.53  \\

    \end{tabular}
    \caption{Number of sources and Median S/N of spectra with coverage of the entire masked broad line (line centre $\pm$10,000\,km\,s$^{-1}$), for the full and S/N > 10 samples.}
    \label{tab:data_sn}
\end{table}
Here we analyse the model's ability to draw from information across the entire unmasked spectrum in order to predict the masked broad lines. All the broad lines in Table~\ref{tab:data_sn} are always masked in testing. The regions constitute $\pm$10,000\,km\,s$^{-1}$ around the rest-frame centre of the line, apart from around \civ\ and \siiv\ where we mask the rest-frame region between 1300--1650\AA\ in order to entirely mask any potential broad absorption line features, which we discuss further in Sec. \ref{sec:comparingtoICA}.

\begin{figure}
    \includegraphics[width=\linewidth, trim = 85 20 105 70, clip]{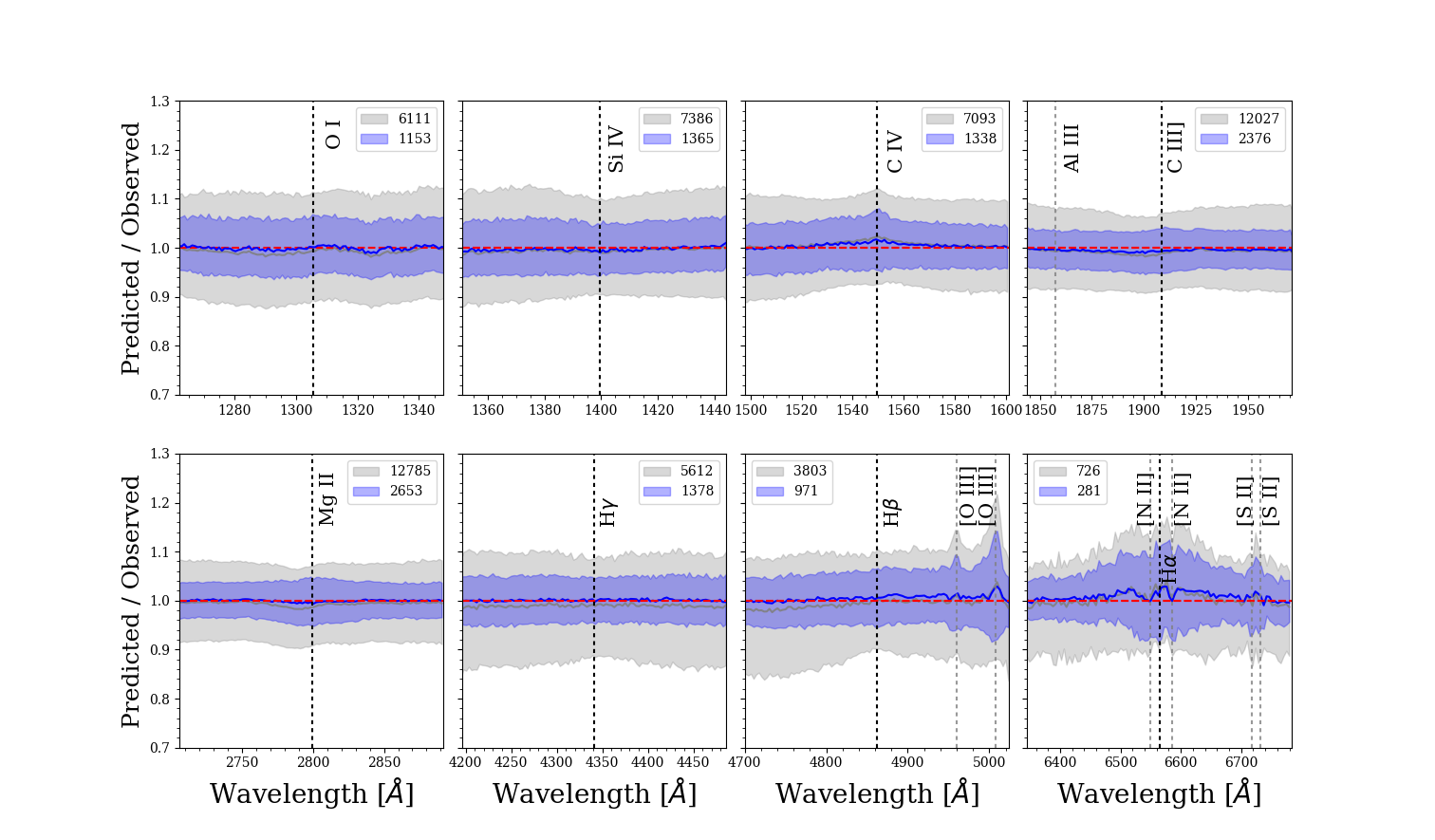}
    \caption{Comparison of predicted/observed flux ratios for several broad lines in the broad line masked test samples, where predictions have been made from the unmasked continuum and weaker lines. The median (solid lines) and 1$\sigma$ ranges (shaded regions) are shown for the full sample in grey and S/N$>$10 sample in blue. The dashed horizontal red line signifies P/O $=$ 1 while dashed vertical lines indicate relevant rest-frame emission wavelengths as labelled. The entirety of these windows has been predicted by the transformer, which each constitute $\pm$10,000\,km\,s$^{-1}$ from the centred broad line. The number of test set spectra that provide full coverage of a particular line are indicated in the legend for each sample. In Appendix~\ref{app:error_floor}, we explore the intrinsic error floor and find that the spreads flattens at 2-6\% at S/N $\approx$ 40 suggesting the error in the data dominates below S/N 40 and error in the model dominates above this S/N.
    } 
    \label{fig:all_broad}
\end{figure}


The left panels of Fig. \ref{fig:singlespec} we show two examples of the broad line masked test case. None of the spectra in our testing have been previously seen by the model.
The QSO spectra from SDSS often contain large noisy regions, which can often contaminate the broad line and likely impact the quality of parameters extracted. We see from these individual cases that the model tends to create a smoothed spectrum and predict through the noise. The model successfully predicts the peaks and widths of each broad line, despite only receiving the AGN continuum and less prominent broad lines, such as the higher order Balmer lines (\hd, \he), He and blended \feii. The results highlight the strongly coupled information that the transformer is able to leverage. In a subsequent paper, we explore the attention mechanism to try and gauge where the model is predominately looking to perform its predictions.

In Fig. \ref{fig:singlespec}, we also show the predicted aleatoric (data) uncertainty, $\sigma$, from the model. In the bottom panel where we show the ratio of predicted/observed (P/O) flux with the predicted uncertainty overlaid, we see that the error ranges are generally comparable to the range in P/O.
These results imply that the model correctly self-predicts flux uncertainties across the spectrum using only the surrounding unmasked fluxes and errors, with no access to the masked ranges. The predicted errors appropriately increase in noisy spectral regions, presumably because the model has learnt during training that wavelengths near the spectral edges are noisier due to a combination of instrumental, sky, and redshift effects. Additionally, despite accurately recovering the shape and peak flux of \ha, the model assigns a large uncertainty to its amplitude and profile, consistent with \ha\ being the most under-represented line in the training set, as reflected in Table~\ref{tab:data_sn} and Fig.~\ref{fig:sn_vs_wav}. The inclusion of the errors of the input spectra allow the model to assign a weight to each flux value, effectively de-prioritising noisy regions of the spectrum, i.e. the edges of the spectrum or telluric lines, and extracting richer information from high-S/N regions.
\begin{figure}
    \includegraphics[width=1.0\linewidth, trim = 70 20 105 70, clip]{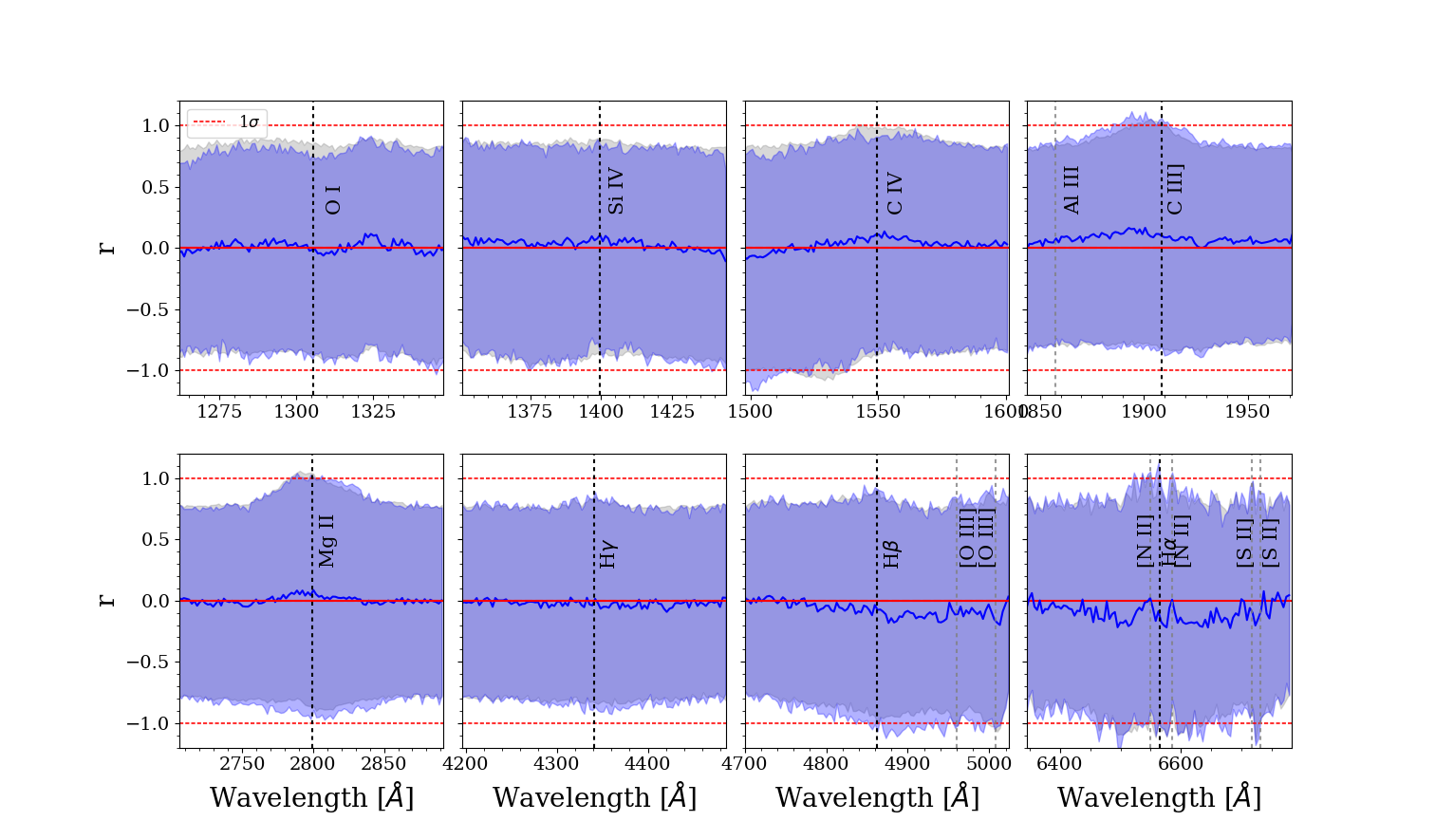}
    \caption{Pull residual ($r$) distribution plots for each of the broad lines shown in Fig. \ref{fig:all_broad} where $r = (f_{obs} - \mu)/\sqrt{\sigma_{pred}^{2} + \sigma_{obs}^{2}}$, with $f_{obs}$ the observed flux value, $\mu$ the predicted mean flux value, $\sigma_{pred}$ the predicted error on the flux value and $\sigma_{obs}$ the quoted error on the flux. A well-calibrated model should follow a standard normal distribution, with a mean of 0 (red solid line) and standard deviation of 1 (red dashed lines).
    } 
    \label{fig:all_broad_err}
\end{figure}
We found that the ratio of predicted/observed fluxes shown in the bottom panels of Fig. \ref{fig:singlespec} does not effectively convey the typical level of deviation across the data. Therefore, we also explore the predictions of the entire test sample in order to build up a statistical understanding of the error within the prediction. We first refer the reader to Table~\ref{tab:data_sn}, which lists the number and median S/N of the spectra that contain full coverage of each broad line that we study, for the full and S/N > 10 samples. These are important to consider when looking at the predictions of the model over the entire test sample, as much of the prediction error will arise from the spectrum noise, as is clear in the \civ\ prediction on the left panel of Fig. \ref{fig:singlespec}.

The results of our broad-line masking test are shown in Fig.~\ref{fig:all_broad}, which compares the predicted/observed flux ratios for \oi, \siiv, \civ, \ciii, \mg, \hg, \hb\ and \ha\ (from left to right, top to bottom) in our test set. The plotting windows cover the $\pm$10,000\,km\,s$^{-1}$ range that is being predicted by \texttt{AGNFormer}. The ratio median (solid lines) and 16th--84th percentile distributions (hereafter $\approx$1$\sigma$; shaded regions) for spectra from both the full (grey) and S/N$>$10 (blue) test sets are shown; the number of spectra used for each subset are listed in the legends.

Overall, we see a consistent decrease in the 1$\sigma$ dispersion of the prediction ratios with increasing S/N across all wavelengths, implying that the dispersion at lower S/N is dominated by spectral noise rather than poor model predictions. We find that this improvement flattens out at $\pm$4\% around S/N $\approx$ 40 (Sect. \ref{app:error_floor}), which we infer to be when the error in the model prediction begins to dominate.

We generally see that the predictions for the full test samples have a dispersion of $\approx$10--15\% while testing on the S/N$>$10 data leads to dispersions of $\approx$4--8\%. The model prediction ratio distributions generally appear relatively symmetric about the 1:1 relation and smooth across most prediction windows, although we do 
note slight overestimations around the peak of \civ\ in the S/N $>$ 10 sample, likely related to residual (mini-BAL) absorption that may have gone undiagnosed by $P_{\rm BAL}$ in some sources, as well as $\approx$1.5--2 times higher median residuals and 1$\sigma$ dispersion in the vicinity of \oiii\ 5008 and the wings of broad \ha, possibly related to the lower numbers of sources used in training; larger datasets improve predictive power, as expected.
Notably, the model is able to predict the masked \oiii\ lines to within a factor of two of the accuracy of the broad lines, highlighting the ability of the model to extract information and create learned connections to narrow lines and continua as well. The larger dispersion associated with the \oiii\ lines may reflect a physically irreducible variance in NLR geometry and filling/cover factors among AGN that is not encoded in the observable spectrum \citep[e.g.,][]{baskin2005b}.
The spreads are smallest for the \ciii\ and \mg\ windows, which likely results from them being the most present lines in the dataset rather than simply being easier to model.

In Fig. \ref{fig:all_broad_err} we also plot the distribution of the residuals with the predicted error factored in, otherwise known as the pull distributions, where $r = (f_{true} - \mu)/(\sigma_{pred}^{2} + \sigma_{true}^{2})^{0.5}$. This plot allows us to gauge how well-calibrated the error prediction is from the model. As we are minimising a Gaussian negative log-likelihood, we are expecting $r$ to be modelled as $\mathcal{N}(0,1)$.
We see that for all broad lines, in both the full and S/N $>$ 10 samples, the distributions lie close to the 1$\sigma$ bounds, especially around the peaks of certain lines like \ciii, \mg, \hb\ and \ha. We interpret Fig.~\ref{fig:all_broad_err} to indicate that the model has successfully calibrated its error estimation assuming Gaussian errors. Near the peaks of the lines, the errors are often fractionally smaller due to the higher flux and can be contaminated by narrow-lines from the NLR or the host galaxy; nonetheless, the model remains well-calibrated here (median $\approx$0; 1$\sigma$ range $\approx$$\pm$1). Moving toward the edges of the windows, the spreads tend to decrease to $\approx$$\pm$0.8, implying that the uncertainties may be overestimated by $\approx$20\% and the model is too conservative in its error assignment. We also see occasional mild asymmetries in the median and 1$\sigma$ distributions at the $\approx$20\% level, around the broad wings of \ha\ and \hb\ and blueward of \civ, implying that some additional systematics (or physics) still remains unmodelled.
We note that the inclusion of the observational errors and error prediction, and the robust calibration of them, has resulted in an improved overall flux estimation by the model, in comparison to a model that does not receive or regress the error.

\begin{figure}
    \centering
    \includegraphics[width=\linewidth,height = 6cm, trim = 30 0 45 50, clip]{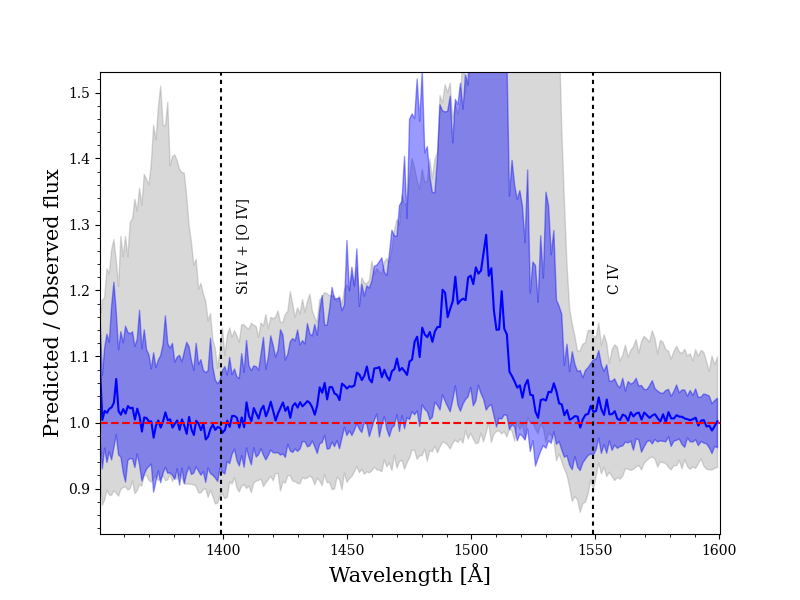}
    \caption{The plot is identical to Fig.~\ref{fig:all_broad}, but displays the combined \civ\ and \siiv\ $+$ \oiv\ range and only shows the predicted/observed flux ratios for the validation sources with $P_{BAL} > 0.95$}. 
    \label{fig:justbal}
\end{figure}
Focussing on the model predictions for \civ, we note that absorption features are sometimes observed near the peak and bluewards of this line; so-called BAL QSOs exhibit the deepest and widest troughs, strongly reshaping the observed line profile. We split out from the test sample potential BAL QSOs using $P_{BAL} > 0.95$ (strong absorption) and show their model predicted/observed ratios in Fig. \ref{fig:justbal} for \civ, as in Fig.~\ref{fig:all_broad}, for both the full and S/N$>$10 samples. Redward of the \civ\ line centre, we find similar dispersion values to the total QSO sample shown in Fig.~\ref{fig:all_broad}, with little deviation as a function of S/N. Blueward, however, we find much stronger dispersion for both S/N subsets. Between $\approx$+3,000-16,000\,km\,s$^{-1}$, the median predicted ratio appears to increase by up to $\approx$30\% above the unity line, while the 1$\sigma$ distributions appear to skew upward by up to $\sim$40-60\% (i.e., overprediction) before levelling out. A similar, but less pronounced trend is seen for \siiv+\oiv.
Without a ground truth to compare to, it is impossible to determine if the model is predicting the observed intrinsic \civ\ flux. Nonetheless, we explore comparisons to algorithms designed especially to reconstruct the line in Sect.~\ref{sec:comparingtoICA}.


\subsection{Half spectrum estimation}
\label{sec:half_spec_res}
The masking method we used in training offers a high level of versatility for model predictions as a function of masking size, range and wavelength. As one specific case, we mask the blue or red half of the spectrum for every source in the test sample, to establish the predictive capabilities of \texttt{AGNFormer} across the broad redshift range of DR16Q.
The central and right portions of Figure~\ref{fig:singlespec} show individual predictions for four sources, where two have the blue half masked and two have the red half masked.
We see in the top panels, \lyman\ and \civ\ have been predicted from the red half of the spectrum, providing impressive estimations of the broad lines and reconstructions in areas of potential broad line absorption (\civ) and IGM absorption (\lyman). We also include the prediction of the error and note that the model assumes larger error in the regions affected by absorption.
In the bottom two panels we predict \oiii\, \ha, \ciii\ and \mg\ from the blue half of the spectrum. The model again provides accurate predictions for the broad lines and continua alike. For \mg\ in the bottom right panel, we see that the model provides a prediction through the strong telluric lines.

An interesting use case for this would be the extension of an observed spectrum to bluer/redder wavelengths than the range covered by a given spectrograph. With enough training across a large redshift range, the model will have an understanding of where and how strongly \ha\ should appear for a high-redshift source or \civ\ for a low redshift source.
This could 1) enable consistent rest-frame comparisons across a wider redshift baseline than currently possible from a given set of instruments, 2) allow better homogenization of datasets across heterogeneous instruments, 3) allow to flag targets likely to show strong features in unobserved windows, informing follow-up prioritization, 4) and providing pseudo-constraints on fluxes in unobserved bands that could help to break degeneracies and improve SED fitting and photo-z estimation.
Unfortunately for the SDSS spectra in our sample, the blue and red ends both suffer from higher noise and therefore generate noisy comparisons, hence why we chose to show only individual cases for this section.

\subsection{Predictions across the QSO MS}
\label{sec:res_qsoms}

\begin{figure}

    \includegraphics[width=1\linewidth, trim = 98 40 75 45, clip]{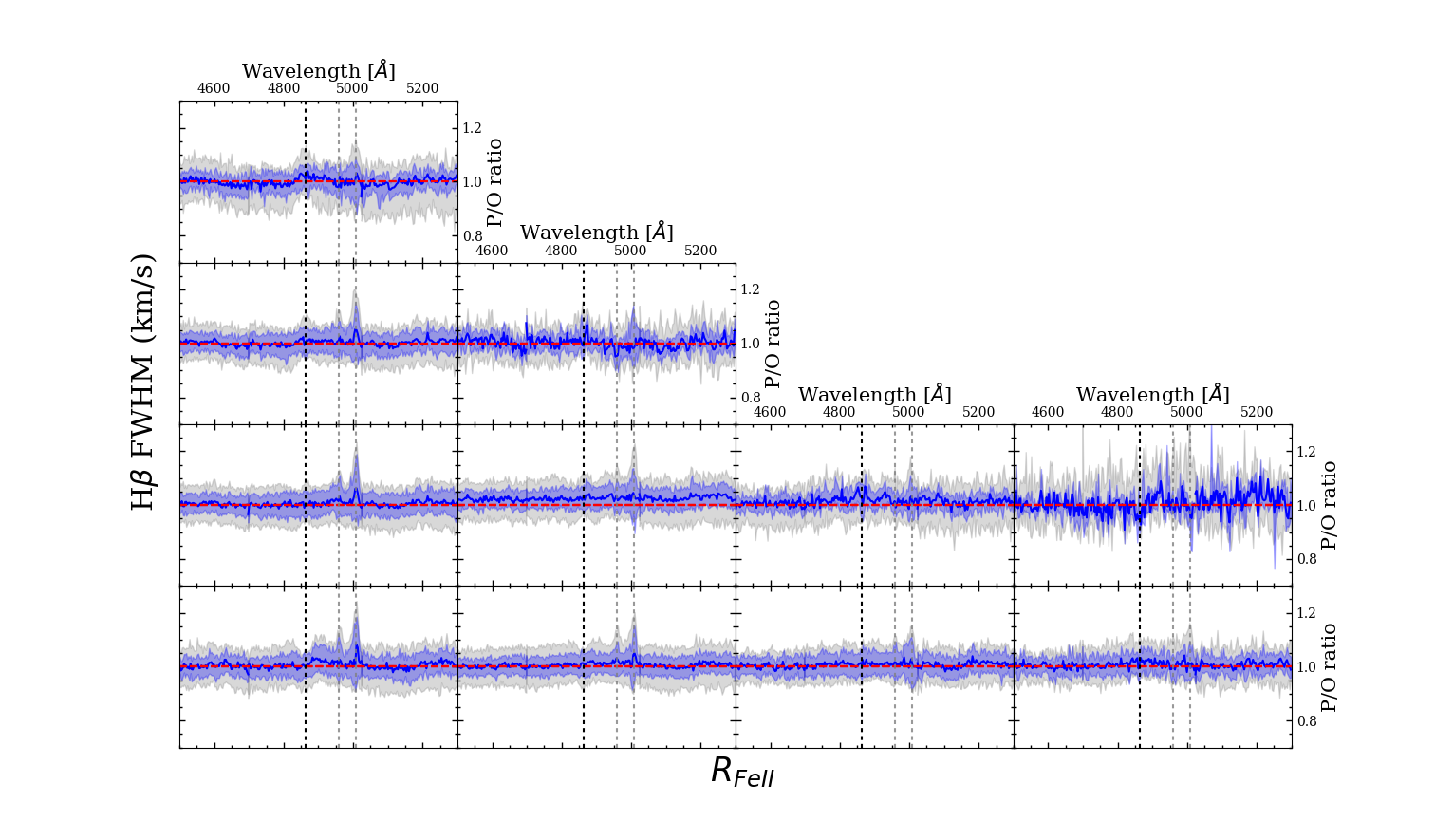}
    \caption{Median predicted/observed ratios (denoted P/O) on the same scale as Fig.\ref{fig:all_broad} for the extended \hb\ region for our test sample, split across bins of \hb\ FWHM and $R_{\rm \feii}$ as highlighted in the QSO MS (top panel of Fig. \ref{fig:density_plots}). We show and predict the \feii\ complexes either side of \hb. Vertical dashed black lines represent the rest-frame \hb\ centre, and the grey dashed lines represent the centres of \oiii$\lambda\lambda$4960,\,5008.}
    \label{fig:optical_qsoms}
\end{figure}

As highlighted in Fig. \ref{fig:density_plots}, there is a large imbalance across the optical and UV QMS, and thus it is important to assess model performance over the entirety of each parameter space.
Early iterations of the model struggled to predict consistently across the entire main sequence. At lower FWHM (v $<$ 8000 km/s), where the majority of the sources lie, the \hb\ line can modelled by a Lorenztian profile, which the predictions reflected. However, at higher FWHM (v $>$ 8000 km/s), the \hb\ profile becomes Gaussian and in these cases early iterations of the model would often still predict a Lorenztian profile, effectively delivering a mean prediction across the full sample, even after attempting sample rebalancing.  We find that the inclusion of random masking, input and predicted errors allow the model to create a strong distinction between a line profile which presents itself infrequently in the dataset with low error (e.g. Gaussian profiles) and features with high error (e.g. artefacts). Essentially, the errors add weights to certain fluxes analogous to performing data augmentation and increasing the likelihood of the model seeing the Gaussian profile, and making consistent predictions across even the low density regions of the QMS. These additions will likely allow the model to be effective at estimating parameters such as \bhm\ across the QMS. This current version of the model performs far better than a mean prediction and is flexible with the line profile that is can predict.

Figure~\ref{fig:optical_qsoms} shows the median predicted/observed spectra as in Fig. \ref{fig:all_broad} for the extended \hb\ region, for each cell across the optical QMS (see Fig. \ref{fig:density_plots}). In this case we have extended the mask to encompass broad \feii, such that the entire 4500\AA -- 5300\AA\ range is predicted.  The broad \feii\ complexes flanking \hb\ are included to demonstrate that the model captures the entire complex structure and connections between these lines and underlying continuum. Predictions are consistent and essentially flat across the full window and the entire QMS, even in lower-density regions where dispersion increases slightly (top left) or becomes noisier (top right). Mildly elevated dispersion near the \oiii\ lines persists, as discussed in Sec.~\ref{sec:bl_results}. This comprehensive understanding of line physics is exactly what is needed for the model to perform well on downstream tasks such as \bhm prediction across a wide dynamic range. The accuracy of the predicted line ratios may yield useful insights into continuum--BLR--NLR connections.

\begin{figure}

    \includegraphics[width=1\linewidth, trim = 100 20 60 40, clip]{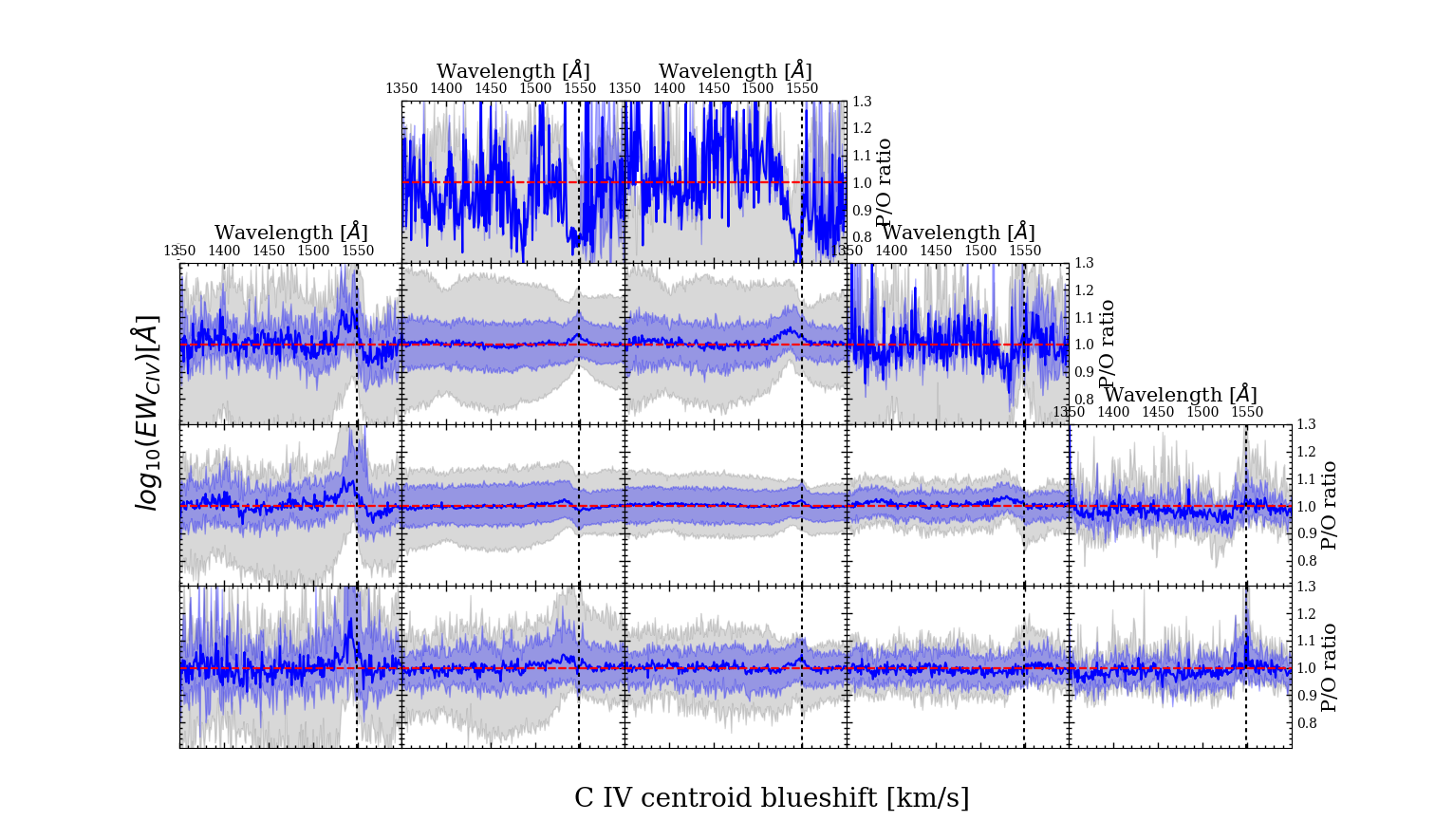}
    \caption{Median predicted/observed ratios (denoted P/O) on the same scale as Fig.~\ref{fig:all_broad} for the \civ\ and \siiv+\oiv\ region for our test sample, split across bins of \civ\ EW and \civ\ offset as highlighted in the UV-QMS (bottom panel of Fig.~\ref{fig:density_plots}). Some of the panels exhibit a lot of noise due to the low density of sources. }
    \label{fig:uv_qsoms}
\end{figure}

Similarly, Fig.~\ref{fig:uv_qsoms} shows the ratios of predicted/observed fluxes split into bins across the UV-QMS (see Fig. \ref{fig:density_plots}). Akin to the changing line profile in Fig. \ref{fig:optical_qsoms},
the blueshift and EW of \civ\ is known to play an important role in tracking the UV-QMS \citep[e.g.,][]{Sulentic2007,richards2011}, with the \civ\ EW further tied to intrinsic luminosity via the Baldwin Effect \citep{baldwin1977}, and the blueshift widely interpreted as tracing accretion disc winds/outflows, whose prominence varies systematically along the sequence \citep{coatman2016,sulentic2017}.
As previously mentioned, the \civ\ line can also exhibit broad absorption troughs related to strong outflows along the line of sight. Initial model iterations were not able to reproduce well the blueshift and skewed profile of \civ\ for velocities $\gtrsim$1500\,km\,s$^{-1}$. Again, the inclusion of randomised masking and error predictions have accelerated the model's ability to learn features in areas of low data density.
Similar to the optical QMS, we see relatively symmetric and flat predictions across much of the windows, for a majority of the UV-QMS panels, although systematic departures occur near and up to $\approx$6,000\,km\,s$^{-1}$ blueward of the \civ\ line core. Specifically, despite the removal of $P_{\rm BAL}$$>$0.7 sources from the comparison for Fig.~\ref{fig:uv_qsoms}, the median and dispersion still show \civ\ overpredictions of up to $\approx$4-6\% in high density regions, and up to $\approx$15-20\% in sparsely populated regions. Conversely, for the full sample, the 1$\sigma$ regions tend to show underprediction asymmetries by $\approx$5-10\%. These deviations tend to show up at lower blueshifts than those seen in Fig.~\ref{fig:justbal}, and may imply either a) the presence (and related uncertainties) of additional low-velocity absorption features or b) uncertainty in the intrinsic amplitudes and profiles of \civ. Some of this is presumably related to the sparsity of spectra in the training sets of particular bins.

\section{Discussion}
\label{sec:discussion}

\subsection{Comparing to QSO reconstruction challenge.}
\label{sec:lyman_reconstruction}

\begin{figure}
   \includegraphics[width=1.01\linewidth, trim = 10 10 10 10, clip]{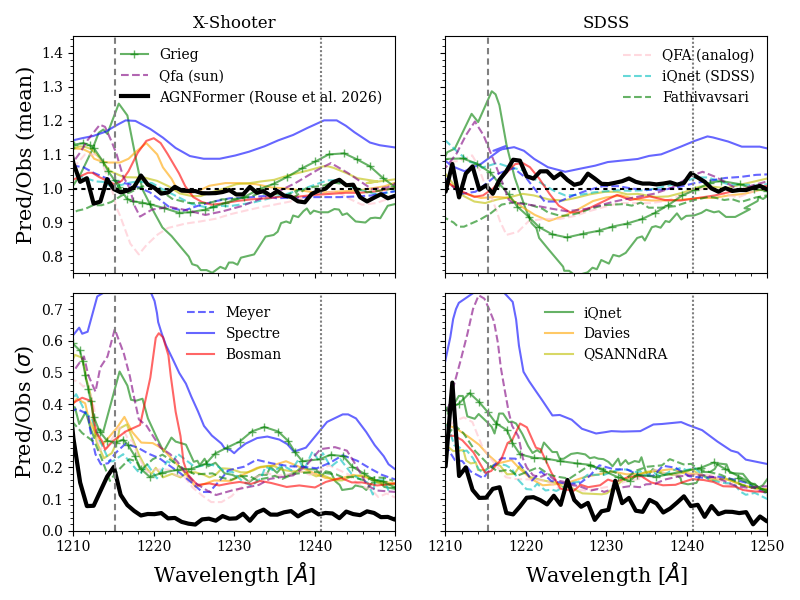}
    \caption{Comparison of the mean (\textit{top}) and absolute error (\textit{bottom}) of the predicted/observed flux ratios from our transformer model (black) vs. 11 models (colours, as indicated among the four legends) from a challenge to predict 30 X-Shooter (\textit{left}) and 30 SDSS (\textit{right}) QSO spectra. This figure is a reproduction of Fig.~8 in \citet{greig2024}.
    }
    \label{fig:lyman_comparison}
\end{figure}

Absorption of far-UV photons at wavelengths in the vicinity and blueward of Ly-$\alpha$ in QSOs, either from intrinsic material (BALs and narrow line absorbers), Ly-$\alpha$ absorption systems along the line of sight, or the neutral IGM at $z\,{\gtrsim}\,6$ \citep[e.g.][]{gunn1965, weymann1981,rauch1998, becker2001}, can strongly hamper measurements of Ly-$\alpha$ and far-UV emission. There are various algorithms, including several ML methods, which leverage spectral information redward of the \lyman\ line in order to reconstruct or predict the intrinsic profile of broadened \lyman\ line and bluer light.

As a means of comparison, \citet{greig2024} conducted a blind challenge to reproduce spectral information between rest-frame 1208--1260\AA\ for 30 SDSS and 30 X-Shooter spectra of QSOs between 3.5 < z < 4.5 (see Sec.~\ref{sec:data}), pitting nine different algorithms (11 models in total after accounting for training variations for two of them) against each other. The challenge excluded wavelengths $\lambda < 1208\AA$ due to the large uncertainties in the amplitudes of the Lyman forest.
The models were trained using a diverse set of spectroscopic data, including objects spanning wide ranges of redshifts and spectral properties, object sample numbers, and input catalogues: SDSS DR12Q \citep{alam2015,paris2017}; DR14Q \citep{paris2018}; and DR16Q \citep{lyke2020}. Among the submitted methods, PCA was the most popular, but not the only method.


To compare our transformer model, we used our random masking model, pretrained on the entire SDSS DR16Q catalogue and then tested using the same 30 SDSS (not present in our training sample) and 30 X-Shooter spectra from the challenge presented by \citet{greig2024}, where the only adaptation we make is to interpolate the X-Shooter spectra in order to reduce the number of spectral pixels to 4500, as this is the sequence length we trained on and hence what the transformer requires. 

In Fig. \ref{fig:lyman_comparison}, we compare the mean (top) and absolute error (bottom) among the predicted/observed flux ratios from our transformer model vs. 11 models from the challenge for the X-Shooter (left) and SDSS (right) QSO spectra, while Table~\ref{tab:qso_reconstruction_values} provides the mean and median absolute errors (MAE) for each spectral dataset and model.
For both samples, the transformer accurately predicts both the peak of \lyman\ line and the overall continuum in the X-Shooter and SDSS samples, outperforming all the purpose built reconstruction algorithms explored in the \citet{greig2024} challenge in terms of mean flux ratio, MAE and median $\sigma$.

An important complementary point here is the fact that our model, trained solely on SDSS spectra (R$\sim$2000), achieves comparable or better performance on X-Shooter data (R$\sim$4500--7500), without any fine-tuning of the model. This resolution invariance highlights the convenience and importance of including a wavelength-based positional encoder to allow the model to deal with the varying wavelength spacing and resolution between the two instruments. This result also mirrors the performance of vision transformers in cross-resolution image tasks \citep[e.g.][]{dosovitskiy2020} and showcases the transfer learning capabilities of our model.

\begin{table}
    \begin{tabular}{l|cc|cc}
    \toprule
    \multirow{2}{*}{\makecell{Reconstruction \\ algorithm}} &
      \multicolumn{2}{c}{\textbf{X-Shooter}}  &
      \multicolumn{2}{c}{\textbf{SDSS}} \\
         &  MAE & Median $\sigma$ &  MAE & Median $\sigma$\\
        \hline
        \midrule
        \texttt{AGNFormer} & 0.017 & 0.052 & 0.022& 0.077\\
        Grieg & 0.061 & 0.266& 0.065 & 0.220\\
        QFA (sun) & 0.067 &0.258&0.058&0.175\\
        QFA (analog) & 0.071 &0.186&0.057&0.169\\
        iQnet (SDSS) & 0.026 &0.207&0.044&0.180\\
        Fathivavsari & 0.025 &0.179&0.058&0.194\\
        Meyer & 0.033 & 0.213&0.025&0.180\\
        Spectre & 0.148 & 0.361&0.100&0.364\\
        Bosman & 0.043&0.237&0.034&0.162\\
        iQnet & 0.125 &0.202&0.130&0.228\\
        Davies & 0.054 &0.219&0.035&0.181\\
        QSANNdRA & 0.046 &0.194&0.023&0.179\\

    \end{tabular}
    \caption{MAE of mean predicted/observed spectra from y = 1. Median value of the 1$\sigma$ distribution of the 30 spectra in each sample.} 
    \label{tab:qso_reconstruction_values}
\end{table}

\subsection{BAL reconstruction}
\label{sec:comparingtoICA}
\begin{figure}
    \centering
    \includegraphics[width=\linewidth,height=7cm, trim = 20 5 50 50, clip]{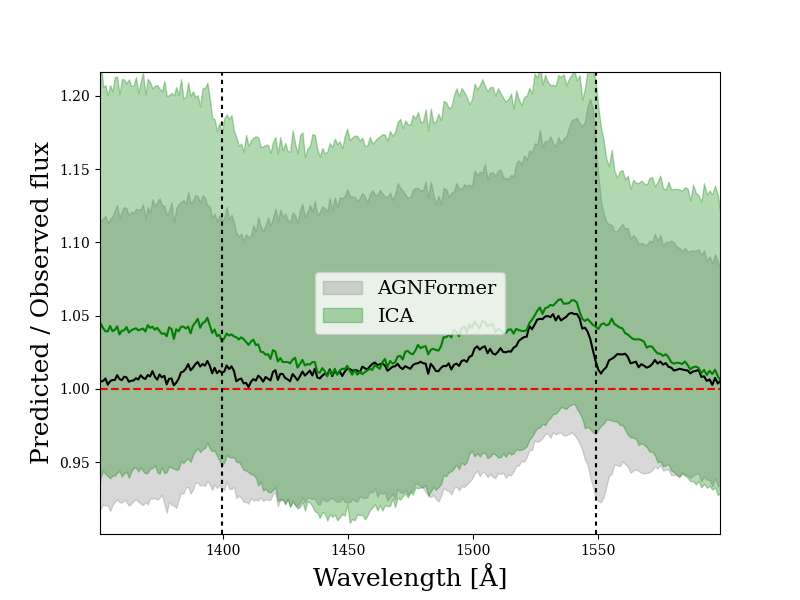}
    \caption{Similar to Fig.~\ref{fig:justbal}, comparing the spectral predictions for the BAL subset of QSOs for the transformer model (black = median, grey = 1$\sigma$ range) and ICA (green = median, light green = 1$\sigma$ range) from \citet{rankine2020}. BALs were excluded from the training of \texttt{AGNFormer}.}
    \label{fig:bal_comp}
\end{figure}

Approximately 15--20\% of optically selected QSOs experience high velocity, non-virialised outflows powered by the central engine, which are observed in the UV spectra as high and low-ionisation BALs \citep[e.g.,][]{gibson2009,rankine2020}. High-ionisation absorption associated with \civ\ is by far the most common \citep[$\approx$85\%; e.g.,][]{tolea2002, gibson2009}.
The absorption is non-virialised, random within some underlying velocity distribution and irrespective of the SMBH mass. As such, it can heavily impact estimations of the intrinsic profile of \civ\ and other lines.

Through early iterations of the model, we found that BALs played an important role in the dataset and affected the model in unexpected ways.
Ideally, if these absorption features are mostly random in velocity and column density and they are fully masked and therefore never seen by the model in testing, the model should reconstruct only the intrinsic broad \civ\ line. However, the model still predicted the presence of absorption, even when masking broad ranges (1300$\AA$-1650$\AA$), effectively drawing on fainter BAL features present in other lines that were seen by the model. Since \civ\ BALs are known to correlate with absorption features across a broad range of ionisation species and wavelengths \citep[e.g.,][]{weymann1991, gibson2009, filiz2014}, we chose to remove BALs from training using the $P_{BAL}$ parameter in \citet{lyke2020}. We retain them in the testing sample, however, allowing us to probe the ability of \texttt{AGNFormer} to reconstruct the \civ\ line in the presence of broad absorption. Figure~\ref{fig:justbal} shows the  predicted/observed flux ratios for the BALQSOs in the validation sample, demonstrating that the model overpredicts the flux blueward of \civ\ in the presence of a BAL. As we do not apply any restrictions to the sources in inference, the model likely receives a substantial number of sources with random absorption features, leading to the very broad prediction excesses seen in the median and 1$\sigma$ distributions.

It is difficult to provide a detailed analysis of the transformer prediction in the absence of ground truth for the \civ\ line. Instead, we compare to models focused entirely on reconstructing the \civ line. Several previous efforts have attempted to reconstruct the intrinsic broad line, in order to allow for a better estimation of its properties (amplitude, FWHM, velocity offset, skew), and by extension those of the absorbers. One of these works, described in \citet{rankine2020}, uses MFICA to reconstruct the \civ\ line, trained on non-BAL QSOs. Here we evaluate the performance of our transformer model compared to that of \citet{rankine2020}.
We stress that aside from removing objects with $P_{BAL} > 0.7$ from training, we make no distinction between non-BAL and BALQSOs.
In contrast, \citet{rankine2020} generate their 10 MFICA components using 4000 fully unmasked non-BAL, spectra, then iteratively mask wavelengths that are directly affected by BAL troughs in each test set spectrum between 1430--1546\AA\ for \civ\ (i.e., blueshifts of $\approx$400--23,000\,km\,s$^{-1}$), until convergence is reached. In testing, the MFICA components are used to deconstruct the masked spectra of BALQSOs and reconstruct predictions over the masked portions of the lines that suffer from absorption.

\begin{figure}

    \hspace{-0.3cm}
    \includegraphics[width=1.05\linewidth, trim = 0 10 0 10, clip]{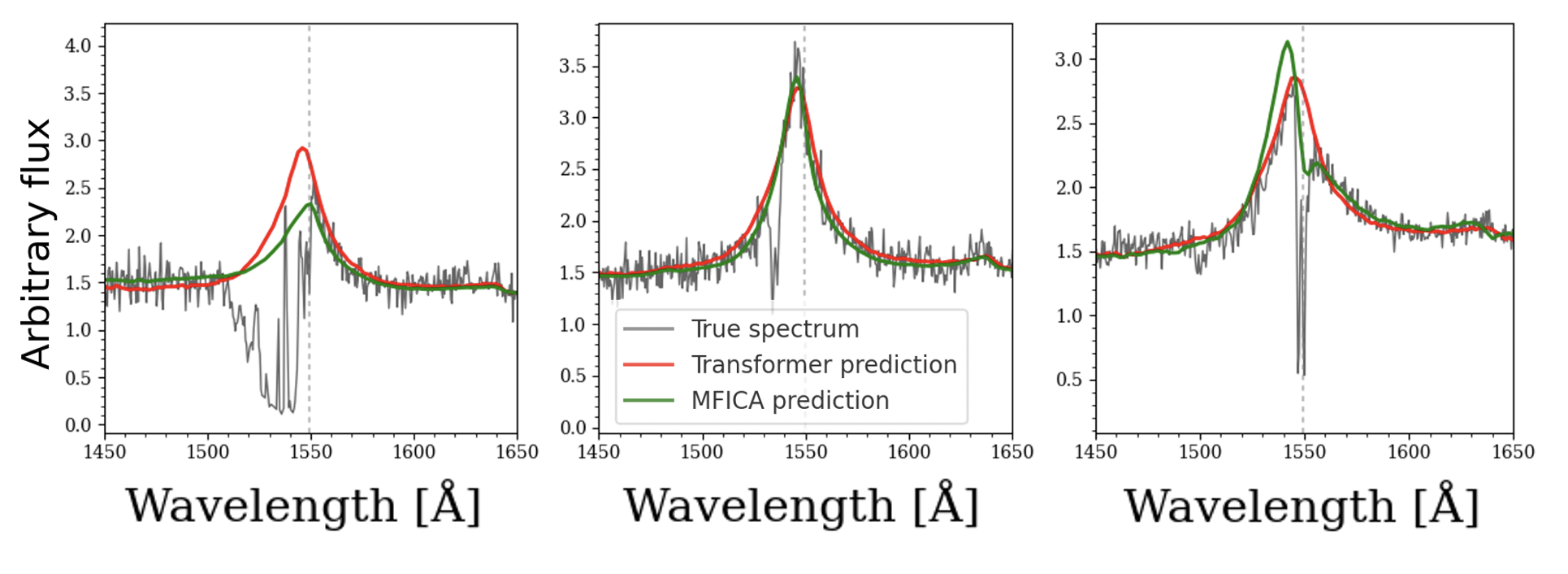}
    \caption{Example predictions around the \civ\ line region from our transformer (red) and the MFICA model of \citet{rankine2020} (green) for different levels and locations of absorption.}
    \label{fig:ICA_comparison_9}
\end{figure}


Comparing reconstruction predictions between the transformer and MCICA for the non-BAL QSOs that overlap with \citet{rankine2020} in Fig. \ref{fig:bal_comp}, we find that the predicted/observed ratios are generally comparable, although the median transformer prediction appears a few percent closer to a ratio of 1.0, while the 1$\sigma$ transformer dispersion appears tighter and more symmetric by up to $\sim$5\%.
In Fig. \ref{fig:ICA_comparison_9}, we display three examples of \civ\ line predictions for BALQSOs, selected from the test sample. There are many cases like the central panel, where both models predict a very similar \civ\ line profile; here the MFICA model appears to predict the amplitude and FWHM more accurately than the transformer. In two other cases (left and right panels), however, the absorption affects the centre of the \civ\ line, leaving more uncertainty in the true line shape. In the left panel, the MFICA reconstruction appears to potentially underestimate the line centre, amplitude, FWHM, and absorption EW, while the transformer predicts a much higher amplitude and slightly blueshifted peak. In the right panel, the MFICA reconstruction clearly overestimates the broad line amplitude and blueshift, and additionally creates a double peaked \civ\ profile, while the transformer appears to generate a reasonably accurate prediction. These differences are due to the adopted choices made to implement the masked regions.
Such results highlight an important future application for the transformer model. In principle further investigation into this topic could be made via analysis of the attention matrix, which we leave as future work.

\section{Conclusions}
\label{sec:conclusions}
This study explores the potential of a pretrained probabilistic transformer-based model to leverage the rich information embedded across the UV and optical spectra of AGN for prediction of masked and unseen spectral regions. We trained the model on a large dataset of spectra from the SDSS DR16Q Catalogue, incorporating the fluxes, flux errors and associated rest-frame wavelengths.

Our model incorporates registers, a custom wavelength positional encoder and a prediction of the mean flux and variance. We find that by including input error and predicted error, \texttt{AGNFormer} performs impressively across the full parameter space of the optical and UV QSO main sequences, and significantly improves model diversity and accuracy over a purely deterministic model. The model provides reliable (16--84 percentile) predictions of masked broad lines to 10--16\% (4--8\%) for the full (S/N $>$ 10) samples. Predictions for unseen (50\% masked) spectra represent well the masked spectrum all the way to the edge of the spectrum. With sufficient data and compute, this could greatly extend the wavelength coverage of spectrographs or flag interesting features in unseen windows for follow-up.

We find that the inclusion of input and predicted error allows the model to build a strong understanding of noisy regions within a spectrum and upweight high S/N regions. With this inclusion, \texttt{AGNFormer} performs better in low data density regimes than a prior deterministic version of the model and can distinguish between a rare, interesting feature and an artefact or noise which will become especially useful when analysing the multiple epochs of spectra from an individual source. We also expect the error inclusion to be advantageous when using this pretrained model in downstream outlier detection tasks.
The pull distributions that we analyse in Fig.~\ref{fig:all_broad_err} show that the model has well-calibrated predictions of its errors compared to a Gaussian distribution of $\mathcal{N}(0, 1)$, although due to the limitations of a Gaussian assumption, it might be slightly overconfident in the error predictions at mask edges.

Our tests demonstrate the model's capacity to accurately learn correlations between various underlying continuum and line emission features, and, importantly, highlight the wealth of information contained in the often discarded or overlooked portions of AGN spectra (i.e., the continuum, blended \feii\ complexes, weaker emission lines). This study unveils new avenues for extracting scientific insights using advanced deep learning techniques. Ultimately, exploring the embeddings formed in this work should enhance our understanding of how fundamental AGN properties, such as SMBH mass, accretion rate, and surrounding structures are encoded within the full AGN spectrum. Having demonstrated that the attention mechanism is capable of leveraging information from the entire spectrum, we will present in a follow-up work (B. L. Rouse et al., in preparation) a self-supervised probabilistic pretraining and supervised down stream probabilistic prediction of AGN properties, including black hole masses.
In a complementary follow-up work (B. L. Rouse et al., in preparation), we will explore the interpretability of the model through its attention matrices. These  allow us to follow the evolution of model-learning, as well as analyse specific lines and spectral regions with respect to their contribution in predicting specific masked broad lines. We provide a Python library with the code used in this work and the pre-trained weights at \url{https://github.com/RouseBen/AGNFormer}.


\begin{acknowledgements}
We kindly thank Bradley Grieg and Amy Rankine for providing comparison samples.
We gratefully acknowledge funding from
ANID grants
CATA-BASAL FB210003 (BLR, FEB, ET),
Millennium Science Initiative AIM23-0001 (FEB, GCV) and NCN2024-112 (GCV), and FONDECYT Regular 1241005 (FEB, ET), 1250821 (ET, FEB), 1231877 (GCV). This research was undertaken with the assistance of resources from the National Computational Infrastructure (NCI Australia), an NCRIS enabled capability supported by the Australian Government, which BLR is very thankful for.
\end{acknowledgements}

%
\bibliographystyle{aa} 
\bibliography{bibliography.bib} 
%
\begin{appendix}
\section{Wavelength-based Positional Encoder}
\label{app:proof}
\subsection{Mathematical Proof}
If the wavelength spacing is uniform, one can prove mathematically that the two positional encoders will produce equivalent frequencies.

From eq.\ref{eq:posenc1}, define $\omega^{(orig)}i \equiv 10000^{-2i/d_{model}}$ so that it becomes: $sin(p\cdot\omega^{(orig)})$.
Then for a uniformly spaced wavelength grid, $\Delta\lambda$, the integer position index $p$ is:
$p = \frac{\lambda-\lambda_{min}}{\Delta\lambda}$.

Thus, if we define:
$\omega^{custom} \equiv \frac{\omega^{orig}i}{\Delta\lambda} = \frac{10000^{-2i/d_{model}}}{\Delta\lambda}$
Then, the original positional encoder can be written as:
\begin{align}
    \text{PE}(\lambda, 2i) &= \sin\!\big[(\lambda - \lambda_{\text{min}})\,\omega^{custom}i\big], \\
    \text{PE}(\lambda, 2i+1) &= \cos\!\big[(\lambda - \lambda_{\text{min}})\,\omega^{custom}i\big],
\end{align}
For $\omega^{(\text{orig})}i = 10000^{-2i/d{\text{model}}}$,
$\omega^{(\text{orig})}{\max} = \omega^{(\text{orig})}{(i=0)} = 1$ and $\omega^{(\text{orig})}{\min}\approx 10^{-4}$. Dividing these by $\Delta\lambda$ gives
$\omega^{(\text{custom})}{\max}=\frac{1}{\Delta\lambda}$ and $\omega^{(\text{custom})}{\min}\approx\frac{10^{-4}}{\Delta\lambda}$.
\subsection{Embedding space}
\begin{figure}
    \centering
    \includegraphics[width=\linewidth]{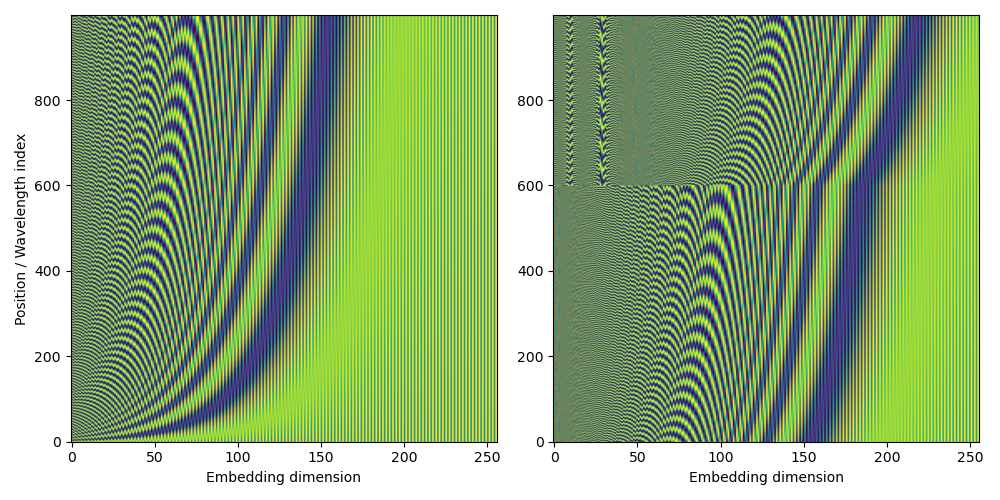}
    \caption{Positional embedding space for toy models with the original positional encoder from \citet{transformer}, described in Eqs.
    2 \& 3 (left) and the wavelength-based positional encoder introduced in this work (right) and described in Eqs. 4 \& 5. An important note is that the left panel will be constant for the original positional encoder, even if $\Delta\lambda \ne const$. Furthermore, the wavelength-based encoder will appear as the left panel where  $\Delta\lambda = const$, but will appear as the right panel when  $\Delta\lambda \ne const$, which is often seen in spectrographs with multiple spectral arms. }
    \label{fig:wavelength_pe}
\end{figure}

In Fig. \ref{fig:wavelength_pe}, we show the positional embedding space for two toy models. The left panel is for a spectrum with constant wavelength spacing ($\Delta\lambda$), which reduces to the same embedding space for the original positional encoder in \citet{transformer}. The right panel presents the case of a spectrum with varying spectral resolution, as might be the case for a spectrograph with multiple spectral arms (split at position index $\approx$ 600).
\section{Model specifics}
\subsection{Batches}
\begin{figure}

    \includegraphics[width=1.\linewidth, trim = 80 10 100 40, clip]{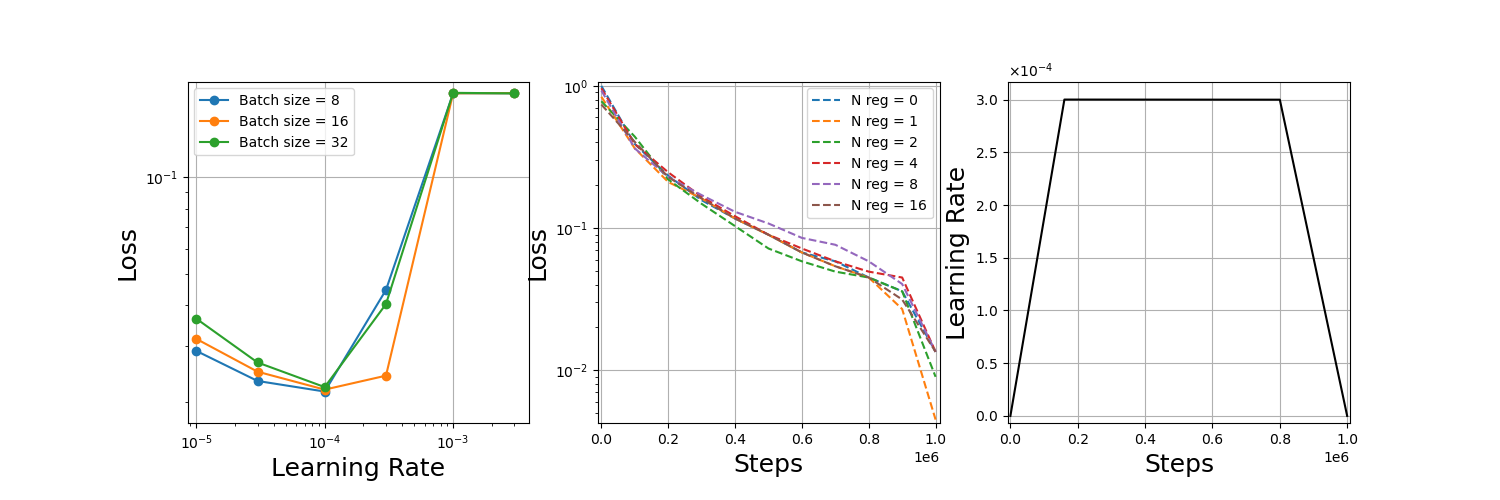}
    \caption{\textit{Left}: Batch size loss minimisation. \textit{Middle}: Register loss minimisation. A comparison of the losses over the entirety of training for each $N_{reg}$ option for $N_{reg} \in\ [0,1,2,4,8,16]$. \textit{Right}: Learning rate evolution over the entirety of training. Formally this learning rate evolution is a warmup-stable-decay learning rate.}
    \label{fig:reg_lr}
\end{figure}

\citet{mccandlish2018} performed an extensive study on the effect of a predetermined batch size on the performance of the model, and searched for an optimal batch size that balanced training speed and efficiency. They note that a larger batch size trains quicker, whereas a smaller batch size is more data efficient. Their Fig. 6 presents the optimisation of steps and examples processed necessary to reach a certain loss, finding that the turnover in efficiency happens around a batch size of 64. The GPU we used\footnote{NVIDIA A100 40GB} allows for a maximum batch size of 32. The left panel of Fig. \ref{fig:reg_lr} shows that the optimal learning rate has the same loss value for the three batch sizes we tested. At their optimal learning rate, the variation between batch size is insignificant. Therefore, we chose a batch size of 32 to allow for faster training.

\subsection{Optimiser/Scheduling}
\label{app:scheduler}
Deep learning models tend to benefit from training with a variable learning rate (the hyperparameter that controls the step size of parameter updates) often adjusted according to a schedule \citep{jacobs1988, schaul2012}. Deep neural networks (DNN) in particular have been shown to be optimisable with decay scheduling \citep{loshchilov2016, touvron2023}. Decay scheduling allows the learning rate to decrease from a maximum to minimum at the end of training. Some versions have used Cosine decays, although, varying training steps and data length require a decay function which can be adapted to suit the needs of the problem.
The warm-up-stable-decay (WSD) or trapezoid scheduler was proposed by \citet{hu2024}, comprised of a linear warm-up from a minimum to maximum learning rate, followed by a stable learning rate over the majority of training and finally a rapid linear decay from the maximum to minimum learning rate. The WSD scheduler has been shown to significantly decrease the loss in LLM training, whilst being adaptable to the desired length of training, considering that the warm-up and decay both happen following a set number of training steps \citep{wen2024, schaipp2025}.

In early iterations of the model, we compared the results from training the model with a static vs. WSD learning rate. The decay to a smaller learning rate at the end of training extracts more performance out of the model than if we were to use a fixed learning rate. Whilst \citet{xiong2020} claim that a learning rate warm-up phase of training is unnecessary due to the pre-normalisation of the transformer, we find that the warm-up and decay phase of training are still useful in our model, regardless of whether we adopt  pre (or post) normalisation. In the right panel of Fig. \ref{fig:reg_lr} we plot the learning rate evolution across training.

\subsection{Model noise bias and error floor}
\label{app:error_floor}
\begin{figure}
    \includegraphics[width=1.0\linewidth, trim = 78 20 105 60, clip]{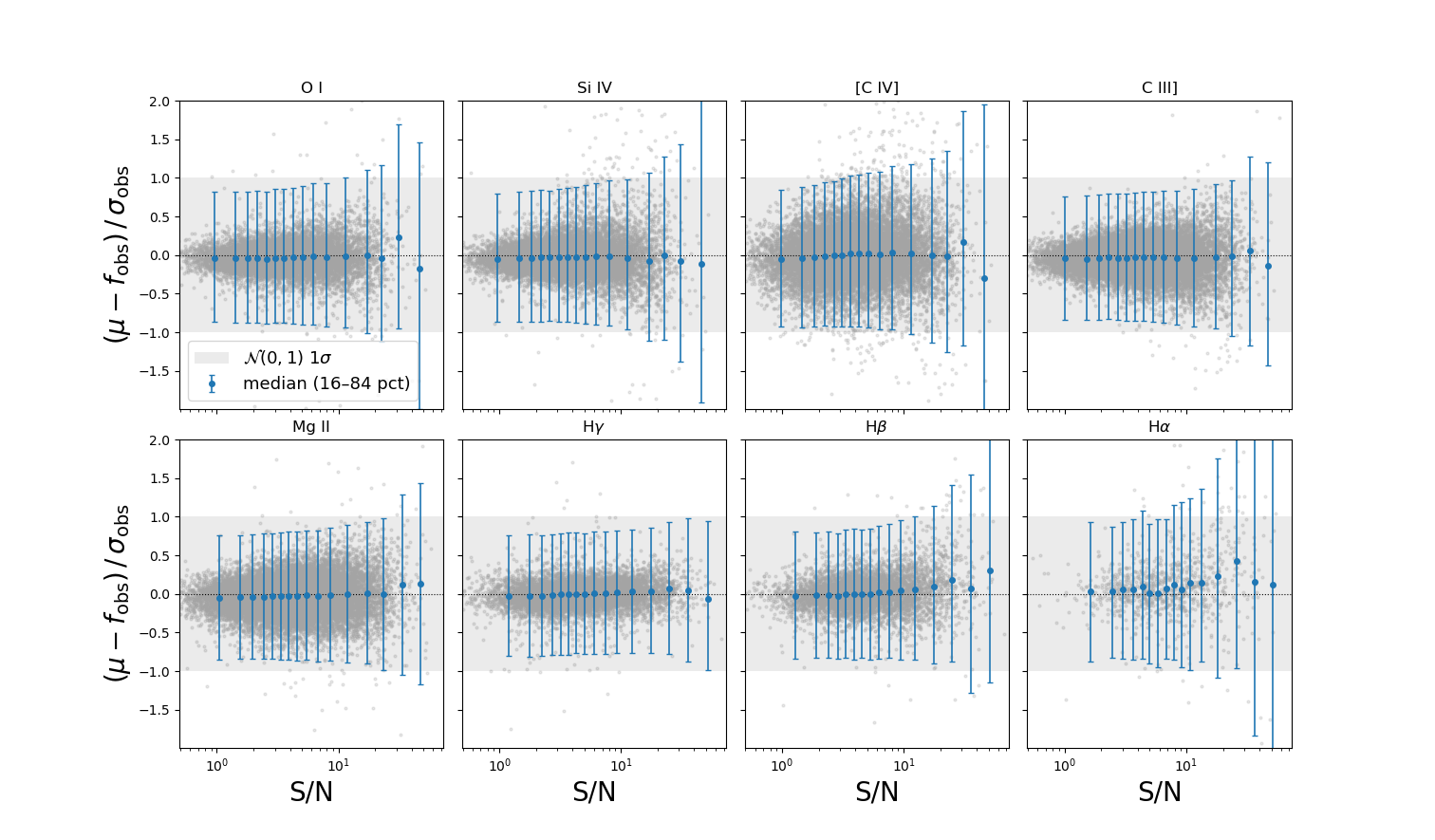}
    \caption{S/N-resolved calibration pull plots for the eight broad line shown in Fig. \ref{fig:all_broad}. To investigate model accuracy, here the pull is calculated using $(\mu - f_{obs})/\sigma_{obs}$ as a function of the median S/N of a spectrum. Error bars denote the median, 16th and 84th percentile for that S/N bin. The grey shaded region highlights a Gaussian centred around 0, with a standard deviation of 1. An accurate model will have a consistent median at 0, error bars within the shaded region indicating model denoising; bars exactly filling the band are consistent with observational noise while bars that extend past the shaded region highlight an error floor. }
    \label{fig:model_pulls}
\end{figure}
Here, we analyse two metrics to determine if there is any noise bias within the model and to define the model's intrinsic error floor. We perform this across each predicted broad line, which allows us to define the error floor per masked region.

First, in Fig. \ref{fig:model_pulls}, we plot the pull distribution of the prediction - observation / $\sigma_{obs}$ as a function of the median S/N of the spectrum. The pull distribution in Fig. \ref{fig:all_broad_err} includes the predicted error and thus measures the calibration of this self-reported uncertainty. The pull distribution here is a diagnostic for the accuracy of the model as a function of the S/N of the spectrum. We see that the median for each broad line is broadly consistent with zero, suggesting that the prediction is unbiased at the majority of S/N. For several lines the median shows a slight negative to positive trend with increasing S/N. This is likely driven by Eddington bias at the faint end, while at the high S/N end it might indicate a multiplicative noise term, although
the lower density of points makes this difficult to quantify. Considering the widths, we find at low S/N that the predictions fall well within the grey Gaussian expectation, meaning they are noise-dominated, although at around $\approx$0.8, this may imply that $\sigma_{\rm obs}$ is overestimated for these lines. At high S/N, the error bars extend past the shaded 1$\sigma$ region, implying an intrinsic reconstruction floor is present.

\begin{figure}
    \includegraphics[width=1.0\linewidth, trim = 78 20 105 60, clip]{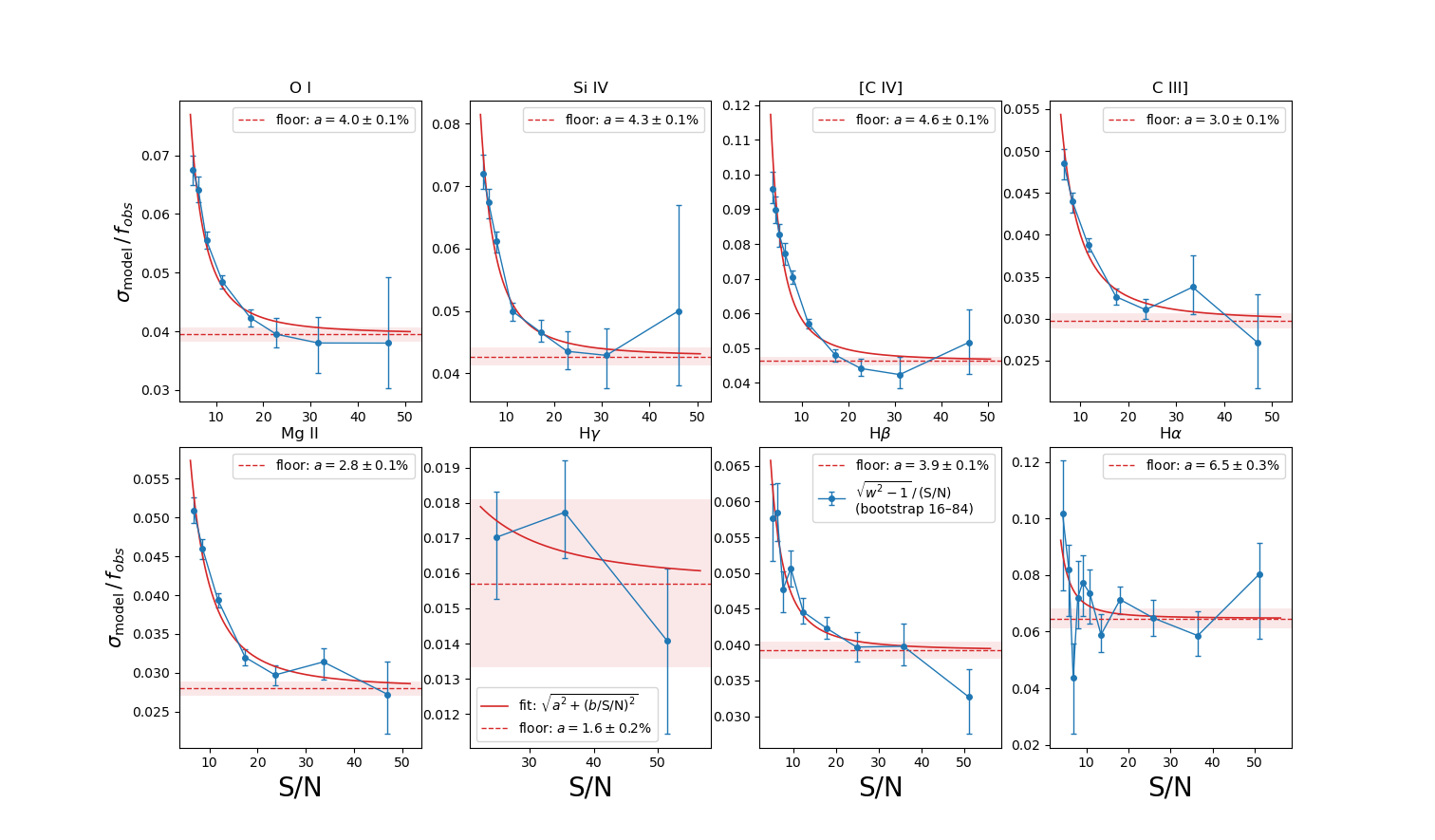}
    \caption{Fractional reconstruction error ($\sigma_{model}$/$f_{pred}$) as a function of the average S/N of a spectrum for each of the masked ranges in Fig. \ref{fig:all_broad}. $\sigma_{model}$/$f_{pred}$ = $\sqrt{w^2 - 1} / (S/N)$ where $w$ is half of the width in the pulls of Fig. \ref{fig:model_pulls}. We fit $\sigma_{model}/f_{pred} = \sqrt{a^2+(b/S/N)^2}$ and identify the asymptote $a$ as the intrinsic error floor for each predicted masked region.}
    \label{fig:frac_unc}
\end{figure}

In Fig.~\ref{fig:frac_unc}, to explicitly define the error floor for each masked region, we search for a plateau in model performance as a function of S/N by plotting the fractional reconstruction error, $\sigma_{obs}/f_{pred} =  \sqrt{w^2 - 1} / (S/N)$, where $w$ is half of the width of the pulls in Fig. \ref{fig:model_pulls}, with uncertainties from bootstrap resampling of sources within each bin.
We find that the fractional error decreases with S/N, as expected if reconstruction quality improves with a stronger signal, we can assume that a model limit has been reached which suggests a genuine floor where more input signal no longer helps.  We fit $\sigma_{model}/f_{pred} = \sqrt{a^2+(b/S/N)^2}$ and identify the asymptote $a$ as the intrinsic error floor, which we quote in each panel. Upward deviations at highest S/N are due to the much lower density of sources in these S/N bins. However, the asymptotes always lie within the error bars of the upturning high S/N points and an upturn never persists over $>$1 bin. The error floors are 4.0$\pm$0.1\%, 4.3$\pm$0.1\%, 4.6$\pm$0.1\%, 3.0$\pm$0.1\%, 2.8$\pm$0.1\%, 1.6$\pm$0.2\%, 3.9$\pm$0.1\% and 6.5$\pm$0.3\% for \oi, \siiv, \civ, \ciii, \mg, \hg, \hb\ and \ha\ respectively.

\section{Telluric line removal}
\label{app:telluric_removal}
As noted in \citet{rankine2020}, their QSO spectra, taken from the BOSS spectrograph, have significant contamination from atmospheric refraction and telluric lines. A detailed discussion on calibration of SDSS-III spectra can be found in \citet{rudolf2016}. Whilst many works exist, which discuss modelling and correction of telluric lines \citep{maiolino1996,vacca2003}, recent efforts in unsupervised ML have proven to also successfully remove telluric lines. For example, \citet{sedahgat2023} use a fully unsupervised deep convolutional auto-encoder to perform telluric line correction on high-resolution stellar spectra (R$=$115,000; 3780--6910\AA) from the High Accuracy Radial velocity Planetary Search \citep[HARPS;][]{mayor2003}.
They serendipitously found that their auto-encoder, despite no specific instruction to remove telluric lines, was able to separate stellar spectra from terrestrial atmospheric lines, and reject the latter. However, they saw a significant drop in the performance when their model was faced with SDSS spectra, which has a much lower spectral resolution of R$\approx$1800.

\begin{figure}
    \includegraphics[width=\linewidth, trim=40 0 60 30, clip]{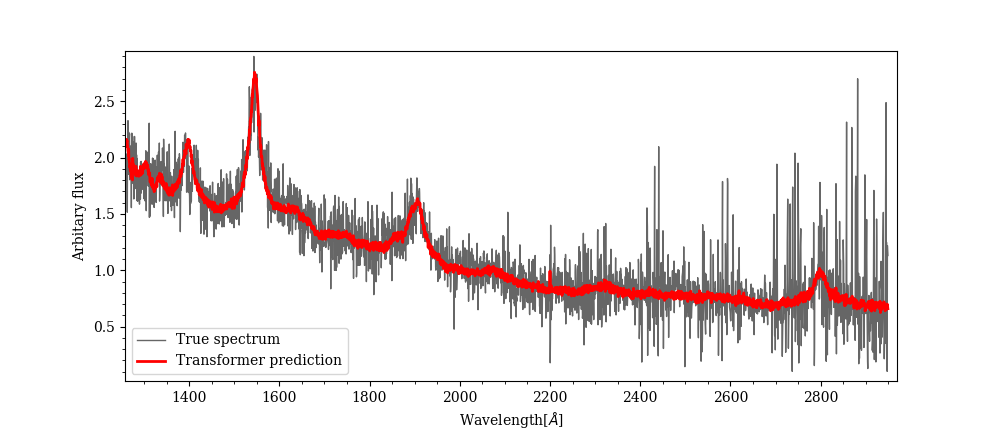}
    \caption{Example low-S/N spectrum strongly contaminated by telluric lines, with much smoother model prediction. When asked to predict all fluxes, rather than a masked range, the transformer model ignores the telluric lines as noise, due to all spectra being shifted to the rest-frame wavelength.   }
    \label{fig:lowsnr}
\end{figure}

We find that the transformer model investigated here also ignores the presence of telluric lines and outputs a much smoother best-fit continuum and emission lines. For this test, we did not mask any of the spectrum \footnote{We also see the model ignore telluric lines in our tests in Sec. \ref{sec:results}.}, meaning the model sees the telluric lines in training, testing and validation. However, because the input spectra are shifted to the rest-frame, the model apparently learns in an unsupervised manner that these lines are contamination and does not predict them. In Fig. \ref{fig:lowsnr}, we show an example spectrum with strong telluric contamination alongside the relatively noiseless model prediction, including a broad \mg\ line in the region with strong  telluric-line contamination.

\section{QMS Comparison Plots}

We show here direct comparisons between median observed and predicted spectra in bins of optical and UV-QMS and median, as complements to Figs.~\ref{fig:optical_qsoms} and ~\ref{fig:uv_qsoms}.

\begin{figure}
    \centering
    \includegraphics[width=0.9\linewidth, trim = 0 0 0 65, clip]{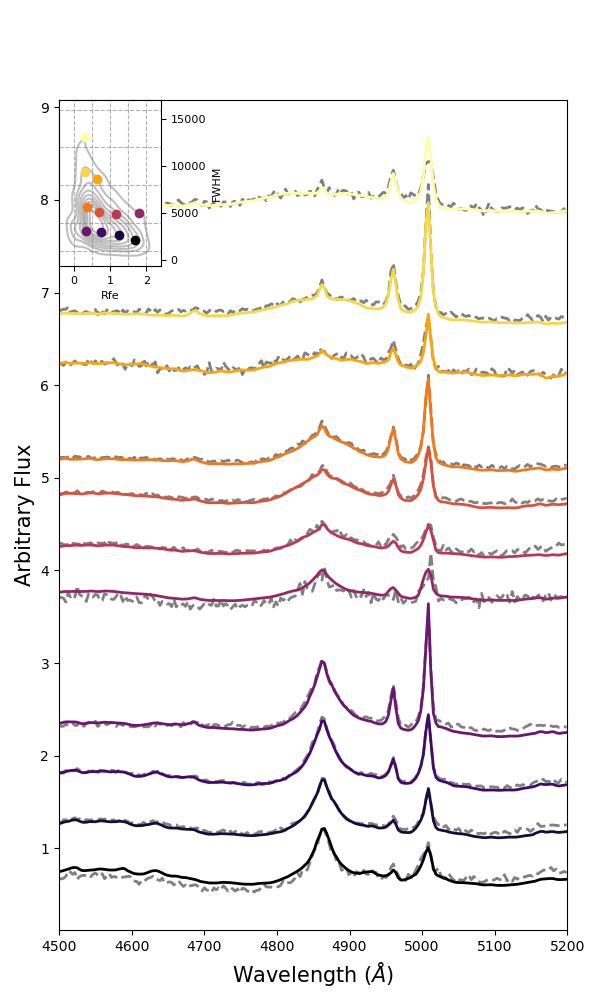}
    \caption{Median observed (grey) and predicted (coloured-coded) spectra for \hb+\oiii+\feii profiles for our test sample, split across bins of optical QMS, as indicated in the top left inset figure, which is a  contour-version of the top panel of Fig.~ \ref{fig:density_plots}. The coloured points in the inset indicate which bin a prediction in the main panel represented.}
    \label{fig:QMS_optcomparison}
\end{figure}
\begin{figure}
    \includegraphics[width=0.9\linewidth, trim = 220 0 100 15, clip
]{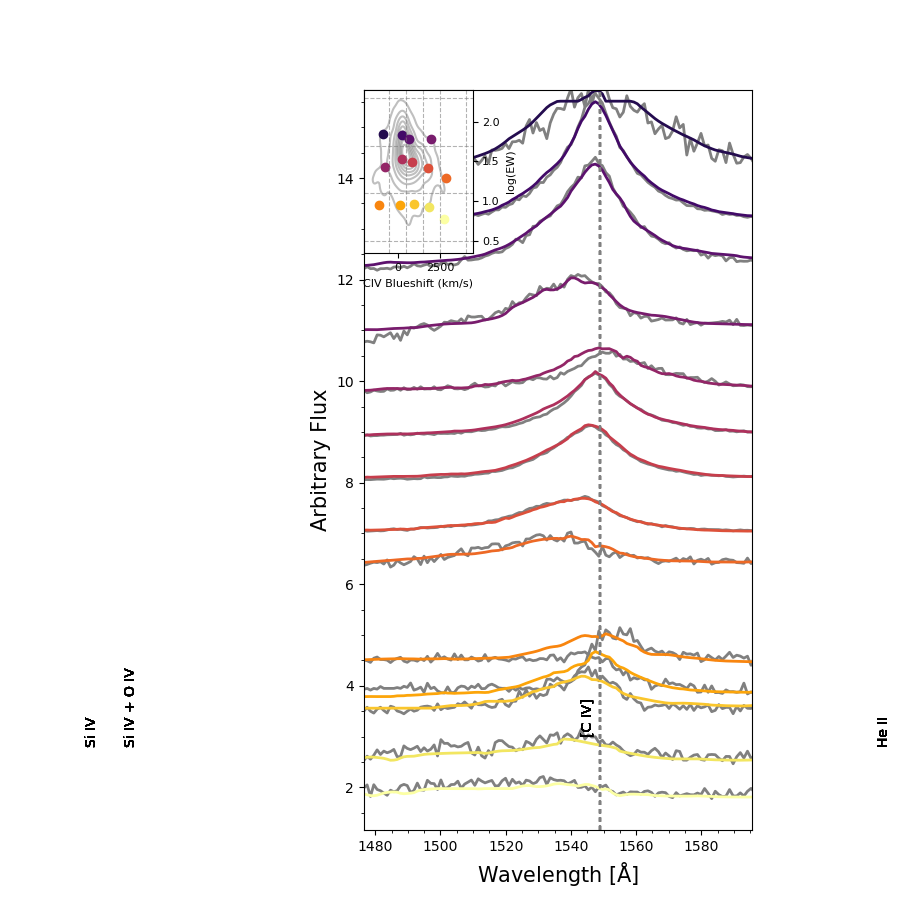}
    \caption{Median observed (grey) and predicted (coloured-coded) spectra for  \siiv+\civ\ profiles for our test sample, split across bins of UV-QMS, as indicated in the top left inset figure, which is a contour-version of the bottom panel of Fig.~ \ref{fig:density_plots}. The coloured points in the inset indicate which bin a prediction in the main panel represented. Dashed vertical lines indicate the rest-frame wavelengths of bright emission lines.}
    \label{fig:QMS_uvcomparison}
\end{figure}

\end{appendix}
\end{document}